\documentclass[journal]{IEEEtran}

\usepackage[
  colorlinks=true,
  linkcolor=blue,
  citecolor=blue,
  urlcolor=blue
]{hyperref}

\usepackage{pgfplots}
\usepackage[linesnumbered,ruled,lined]{algorithm2e}
\usetikzlibrary{shapes.multipart,intersections}
\usepackage{amsmath,amssymb,amsfonts,amsthm,steinmetz}
\usepackage{mathrsfs}  
\usepackage{textcomp}
\usepackage{acronym}
\usepackage{xcolor}
\usepackage{upgreek,xspace}
\usepackage{array}
\usepackage{tikz}
\usetikzlibrary{calc}
\makeatletter
\newcommand{\gettikzxy}[3]{%
  \tikz@scan@one@point\pgfutil@firstofone#1\relax
  \edef#2{\the\pgf@x}%
  \edef#3{\the\pgf@y}%
}
\makeatother

\usepackage{comment}

\usepackage[draft]{todonotes}   
\usepackage[T1]{fontenc}

\usepackage{esvect}

\usepackage{times}
\usepackage{bm}
\usepackage{stmaryrd}
\usepackage{babel}
\usepackage{graphics, graphicx}
\usepackage{gensymb}
\usepackage{cite}
\usepackage{enumitem}
\usepackage{url}

\usepackage{mathtools}

\usepackage[all=normal,paragraphs=normal,floats=normal,mathspacing=normal,wordspacing=tight,charwidths=tight,mathdisplays=normal,leading=tight]{savetrees}

\begin{document}

\title{ADC-Aware End-to-End Optimization of a Dynamic Metasurface Antenna with Strong Mutual Coupling for Monostatic Scene Classification}

\author{Philipp~del~Hougne,~\IEEEmembership{Member,~IEEE}
\thanks{
P.~del~Hougne is with the Department of Electronics and Nanoengineering, Aalto University, 00076 Espoo, Finland and with Univ Rennes, CNRS, IETR - UMR 6164, F-35000, Rennes, France. (e-mail: philipp.del-hougne@univ-rennes.fr)
}
\thanks{This work was supported in part by the Nokia Foundation (project 20260028), the ANR France 2030 program (project ANR-22-PEFT-0005), the ANR PRCI program (project ANR-22-CE93-0010), the European Union's European Regional Development Fund, and the French region of Brittany and Rennes Métropole through the contrats de plan État-Région program (projects ``SOPHIE/STIC \& Ondes'' and ``CyMoCoD'').}
}

\maketitle

\begin{abstract}
Dynamic metasurface antennas (DMAs) enable programmable wave-domain signal processing that can be jointly optimized with downstream digital processing in an end-to-end manner. Existing studies, however, typically assume ideal analog-to-digital conversion (ADC) and often rely on simplified electromagnetic models. Here, we study ADC-aware end-to-end optimization of a monostatic sensing pipeline based on a DMA with strong mutual coupling (MC). We model the wave domain using an MC-aware multiport-network model whose parameters were experimentally estimated for a fabricated chaotic-cavity-backed DMA with 96 one-bit-programmable meta-elements. 
We perform ADC-aware end-to-end optimization of the DMA configurations and digital classifier, either with awareness of a fixed uniform ADC or, optionally, with jointly learned ADC decision thresholds, and compare against baselines that assume an ideal ADC and/or ignore MC. Our results show that ADC awareness is essential in low-resolution ADC regimes: with one-bit ADCs and eight DMA configurations, deploying an ideal-ADC-trained system with a uniform one-bit ADC reduces the test accuracy from $95.5\%$ to $56.0\%$, whereas ADC-aware training with the same fixed uniform one-bit ADC achieves $87.2\%$. We also show that without MC awareness the accuracy drops to the random-guess level. Learning non-uniform ADC thresholds provides at most modest additional gains over fixed uniform ADCs in the considered DMA-based sensing pipeline.
\end{abstract}

\begin{IEEEkeywords}
Analog-to-digital conversion, dynamic metasurface antennas, end-to-end optimization, mutual coupling, task-aware sensing, wave-domain computing.
\end{IEEEkeywords}

\section{Introduction}
\label{sec_introduction}

Next-generation wireless systems for sensing and communications are expected to partially offload their signal-processing burden from conventional digital processors to the wave domain, where electromagnetic signals can be processed in their native domain through reconfigurable wave-matter interactions~\cite{pWDCperspective}. Potential benefits may include improvements in power consumption, speed, and cost. One emerging technological enabler of programmable wave-matter interactions is the dynamic metasurface antenna (DMA). A DMA is an ultrathin structure that couples a small number of feed ports to a large number of radiating elements using waveguides~\cite{sleasman2015dynamic} or cavities~\cite{sleasman2020implementation}; the DMA is parametrized by a set of tunable elements which are typically collocated with the radiating elements (beyond-diagonal DMAs feature additional tunable elements in the waveguides or cavities~\cite{prod2025beyond}). Compared to traditional hybrid antenna architectures consisting of a reconfigurable combining board and an antenna array~\cite{gong2020rf}, a DMA integrates reconfigurable combining and radiating elements in a very compact form factor~\cite{DMA2020}. 

DMAs were initially explored for applications in \textit{compressive sensing}, by multiplexing the electromagnetic signals across a known series of \textit{random} DMA configurations~\cite{sleasman2015dynamic,sleasman2020implementation}. Since often only a portion of the information carried by the electromagnetic signals is useful for downstream processing tasks, applications in \textit{learned sensing} later explored the use of a series of task-aware \textit{learned} DMA configurations~\cite{del2020learned}. The latter were identified through a joint end-to-end optimization of the DMA configurations and the downstream digital processing~\cite{del2020learned}. Learned DMA configurations substantially reduced the number of required measurements in a prototypical scene-classification task~\cite{del2020learned}. Subsequently,~\cite{qian2022noise} endowed the learned DMA configurations additionally with an awareness of the measurement-noise distribution. All these DMA-empowered sensing applications can be interpreted as integrating wave-domain computing and sensing; in the case of \textit{learned sensing}, the wave-domain computation is task-aware such that it pre-selects salient information.

The joint optimization in~\cite{del2020learned,qian2022noise} of information processing in the DMA's wave domain and in the downstream digital domain assumes that there is no information loss at the interface between the two domains. Yet, practical implementations require analog-to-digital conversion (ADC) that quantizes the analog signal and thus inherently leads to a loss of information. 
Of course, high-resolution ADCs could in principle minimize this information loss, but in practice they would imply substantial power consumption and cost~\cite{Zhang2018COMMAG}; moreover, a large acquired bit budget increases the amount of data that must be transferred, buffered, stored, and processed downstream.
Meanwhile, there has been significant progress in ADC-aware signal processing for low-resolution signal acquisition~\cite{kipnis2018analog,Zhang2018COMMAG}.
A first step is to account for finite-resolution \textit{uniform} ADCs when jointly designing analog combining and digital processing, possibly including some ADC parameters such as the ADC's dynamic range in the optimization~\cite{shlezinger2019hardware,shlezinger2019asymptotic,neuhaus2021task,wang2021dynamic,lou2026dynamic}. A further step is to learn the ADC mapping itself: deep task-based ADCs learn task-adapted \textit{non-uniform} ADC quantization mappings jointly with analog combining and digital processing~\cite{shlezinger2022deep,vol2025learning}. 
Yet, to the best of our knowledge, ADC-aware DMA-empowered sensing has not been studied to date, neither with uniform nor with task-aware non-uniform ADC quantization mappings.

While~\cite{wang2021dynamic,lou2026dynamic} consider DMA-based analog combining, their system models do not account for important electromagnetic phenomena such as mutual coupling (MC) within the DMA. Theoretical works often neglect MC as a practical nuisance, but it was recently shown that strong MC boosts the DMA's wave-domain flexibility~\cite{prod2025mutual,prod2025benefits}. Since maximal wave-domain flexibility is naturally desirable, the use of MC-aware system models is important. The experimental works in~\cite{sleasman2015dynamic,sleasman2020implementation} account for MC by measuring the DMA's radiation pattern for the selected random sequence of DMA configurations. However, while MC-aware, this model-agnostic approach is not amenable to optimizing the DMA configurations. The theoretical works in~\cite{del2020learned,qian2022noise} use an MC-aware coupled-dipole model with practically plausible parameters. An alternative MC-aware model has been formulated in terms of multiport-network theory (MNT)~\cite{williams2022electromagnetic}. While the MNT-based model's parameters can be extracted in numerical full-wave simulations~\cite{almunif2025network,ramirez2025metasurface}, it is unlikely that this route is viable for applications to experimental DMA prototypes. \textit{First}, the DMA design may be proprietary and thus unknown to the wireless practitioner. \textit{Second}, if the fabricated DMA design is perfectly known, an accurate simulation of such an electrically large structure can involve a prohibitive computational cost. \textit{Third}, fabrication inaccuracies would likely result in substantial deviations between DMA design and DMA prototype; the sensitivity to fabrication inaccuracies would be particularly high for DMAs with strong MC. Altogether, an indirect estimation of a set of proxy MNT parameters based on experimental measurements with the considered DMA prototype, as recently demonstrated in~\cite{tapie2025experimental}, is required to obtain an accurate MNT-based system model. The attribute ``proxy'' emphasizes inevitable ambiguities in the estimated parameters that do not affect physically measurable quantities (see Appendix~A in~\cite{tapie2025experimental}).

\begin{figure*}
    \centering
    \includegraphics[width=\textwidth]{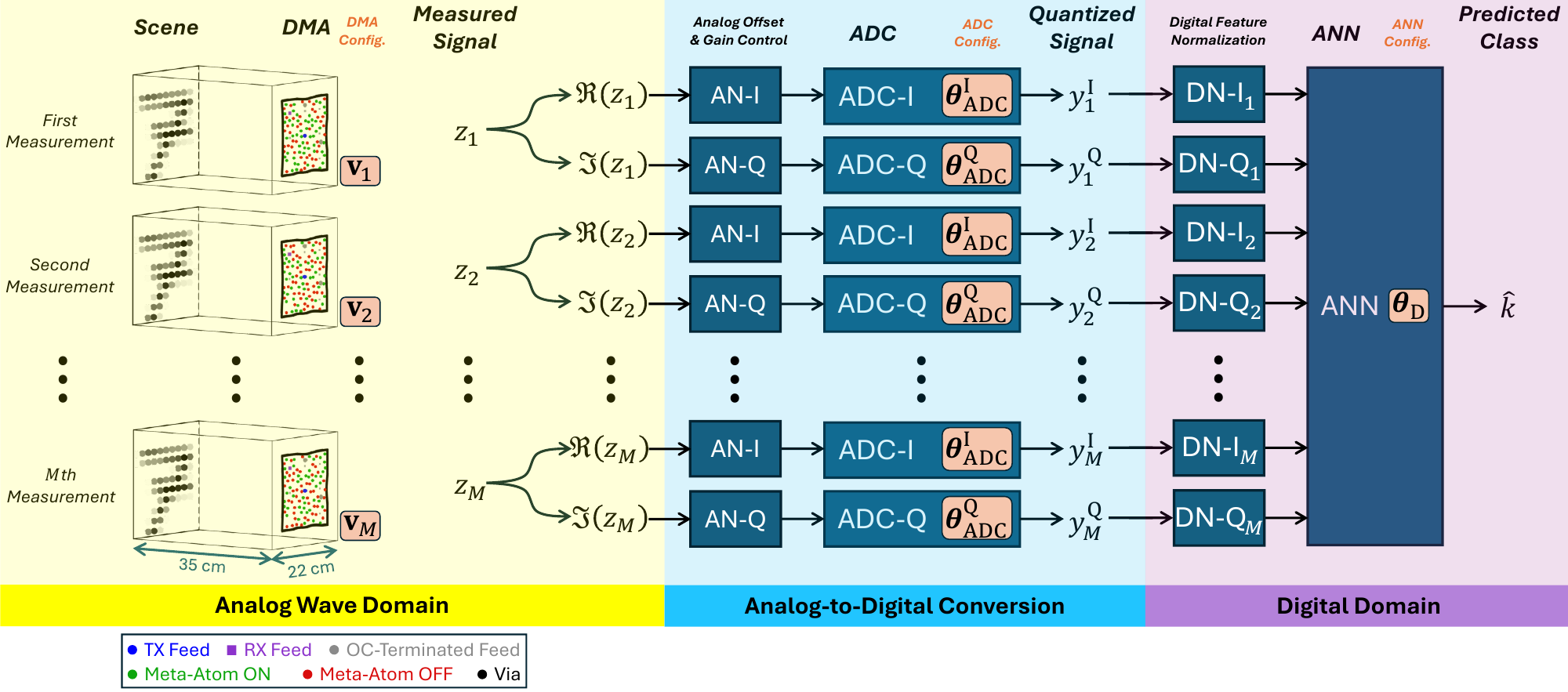}
    \caption{Schematic overview of the considered end-to-end scene-classification sensing pipeline, comprising the analog wave domain, the analog-to-digital conversion, and the digital domain. Each domain features optimizable parameters. Our multi-shot single-input-single-output approach relies on $M$ shared-aperture monostatic measurements with one transmit feed and one distinct receive feed. The two feeds are hosted by a shared DMA comprising 96 1-bit-programmable meta-elements; the DMA configuration differs for each measurement. Unused DMA feeds are terminated with open circuits; further details on the DMA are provided in Sec.~\ref{subsec_DMA_Setup}. The analog signals are offset and gain controlled before ADC. The ADC outputs are normalized in the digital domain before entering the ANN. }
    \label{Fig1}
\end{figure*}

In this paper, we study the ADC-aware optimization of a DMA with strong MC for monostatic scene classification. We use the experimentally estimated proxy MNT parameters from~\cite{tapie2025experimental} to ground our work in a fabricated DMA prototype. 
We deliberately use a single transmit feed and a single receive feed on the same DMA aperture to avoid digital beamforming, thereby minimizing the RF-chain requirements. The shared-aperture monostatic architecture is compact and convenient to measure the received signal with a well-defined phase reference relative to the transmitted signal. The resulting end-to-end sensing pipeline is summarized schematically in Fig.~\ref{Fig1}.
Compared to~\cite{del2020learned,qian2022noise}, the main differences are, \textit{first}, the ADC awareness, and, \textit{second}, the experimental grounding. In addition, we consider a monostatic scenario whereas~\cite{del2020learned,qian2022noise} studied a bistatic setup. Moreover, we use an MNT-based DMA model whereas~\cite{del2020learned,qian2022noise} used a coupled-dipole-based DMA model.
Compared to~\cite{shlezinger2022deep,vol2025learning}, the main difference is that we consider DMA-based analog combining with an MC-aware and experimentally calibrated system model (accounting for the 1-bit-programmability constraint on the DMA's tunable elements), instead of modeling the analog combining with a simple unconstrained real-valued matrix multiplication. 
Moreover, our learned-ADC case adapts only the ADC thresholds shared across the multi-shot measurements, whereas~\cite{shlezinger2022deep} considers a broader task-based acquisition design that also includes sampling choices, and~\cite{vol2025learning} focuses on trainable memristive ADC hardware and its power--accuracy trade-off.
Compared to~\cite{wang2021dynamic,lou2026dynamic}, the main differences are, \textit{first}, the use of a physics-consistent (and experimentally calibrated) MNT model to describe the DMA (accounting for 1-bit-programmable DMA elements), and, \textit{second}, the end-to-end joint optimization of DMA configurations, ADC quantization mapping, and digital processing layer. 
Our main contributions are summarized as follows:
\begin{enumerate}
    \item We report the first end-to-end optimization of a DMA-empowered sensing scheme that is both MC-aware and ADC-aware. We consider a single transmitter and a single receiver sharing the same DMA aperture.
    \item We systematically compare the performance of (i) ADC-aware hybrid analog-digital optimization and (ii) tri-hybrid analog-ADC-digital optimization, as a function of the number of measurements, the ADC bit depth, and the pre-ADC measurement noise level.
    \item We inspect the resulting signal distributions and corresponding optimized ADC thresholds.
    \item We benchmark our results against cases without MC awareness and/or without ADC awareness.
\end{enumerate}
Our systematic comparisons and benchmarking allow wireless practitioners to assess the marginal performance gains enabled by tri-hybrid analog-ADC-digital optimization compared to ADC-aware hybrid analog-digital optimization, and by MC-aware and/or ADC-aware optimization.

Our paper is organized as follows.
In Sec.~\ref{sec_RelatedLiterature}, we discuss additional related literature.
In Sec.~\ref{sec_SystemModel}, we introduce the end-to-end sensing model.
In Sec.~\ref{sec_OptAlg}, we describe the considered optimization cases, differentiable relaxations, and training procedure.
In Sec.~\ref{sec_ExpResults}, we present the experimental setup and scene-classification results.
We close with conclusions in Sec.~\ref{sec_Conclusion}.

\textit{Notation:}
$\mathbb{R}$, $\mathbb{C}$, and $\mathbb{B}\triangleq\{0,1\}$ denote the sets of real, complex, and binary numbers, respectively.
$\Re\{\cdot\}$ and $\Im\{\cdot\}$ denote the real and imaginary parts, respectively.
$(\cdot)^\top$ denotes transpose.
$|\cdot|$ denotes absolute value for scalars and cardinality for sets.
$\mathbf{I}_a$ denotes the $a\times a$ identity matrix.
$\mathbf 0$ and $\mathbf 1$ denote the all-zeros and all-ones vectors/matrices of appropriate sizes.
$\mathrm{diag}(\mathbf{a})$ denotes the diagonal matrix whose diagonal entries are given by the vector $\mathbf{a}$.
$\mathbf{A}_{\mathcal{B}\mathcal{C}}$ denotes the block of $\mathbf{A}$ selected by row indices $\mathcal{B}$ and column indices $\mathcal{C}$.
$\odot$ and $\oslash$ denote elementwise multiplication and division, respectively.

\section{Related Literature}
\label{sec_RelatedLiterature}

In this section, we complement the contextualization of our work provided in the introduction with a broader discussion of related literature beyond DMA-empowered architectures and low-resolution ADC-aware optimization. Indeed, the transition from compressive sensing with pseudo-random configurations to learned sensing with end-to-end optimized task-aware configurations has occurred broadly across many hardware modalities~\cite{saigre2022intelligent}. 

Closest to the DMA-empowered sensing architectures are sensing architectures empowered by reconfigurable intelligent surfaces (RISs). The main difference between a RIS and a DMA is that the RIS does not have integrated feeds, which means that RIS-based architectures require a separate feed antenna such that the overall setup is usually less compact. The MNT models have the same mathematical structure in both cases.
Compressive sensing with a transmissive RIS was demonstrated in~\cite{li2016transmission} while learned sensing with a reflective RIS was demonstrated in~\cite{li2020intelligent}. Neither~\cite{li2016transmission} nor~\cite{li2020intelligent} use experimentally calibrated MNT models;~\cite{li2016transmission} is based on a model-agnostic approach while~\cite{li2020intelligent} uses a data-driven system model that can implicitly account for MC. Both of these works tacitly assume ideal ADC without any ADC-related information loss.

The literature further contains intermediate strategies between pseudo-random configurations and fully task-aware learned configurations~\cite{saigre2022intelligent}. One line of work minimizes redundancy between measurement modes by optimizing their diversity, for instance by reducing correlations between sensing vectors, flattening the singular-value spectrum of the sensing matrix, or maximizing information-capacity-inspired objectives~\cite{obermeier2017model,obermeier2019sensing,molaei2019digitized,del2020optimal,obermeier2020generalized,tapie2026optimizing}. Another intermediate strategy is to additionally exploit prior knowledge about the scene distribution by using principal scene components as measurement modes~\cite{liang2015reconfigurable,li2019machine}. Measurement modes based on principal scene components are (ideally) orthogonal and scene-aware, but they remain task-unaware. All of these works tacitly assume ideal ADC without any ADC-related information loss.

The broader trend toward end-to-end optimized systems is not limited to microwave sensing: optical systems featuring jointly optimized coded illumination, diffractive optics, metasurfaces, or microscope hardware together with downstream reconstruction or classification digital neural networks have been studied~\cite{chakrabarti2016learning,horstmeyer2017convolutional,kellman2019physics,sitzmann2018end,chang2018hybrid,muthumbi2019learned,lin2021end,tseng2021neural,arya2024end}, and analogous ideas have been explored for learned ultrasound beamforming~\cite{vedula2019learning}. These works further support the general principle that task-aware acquisition can reduce the burden on downstream digital processing by acquiring measurements in a task-aware manner. However, all of these works in optics and acoustics tacitly assume ideal ADC without any ADC-related information loss.

\section{System Model}
\label{sec_SystemModel}

In this section, we describe the end-to-end system model for DMA-empowered, ADC-aware, and MC-aware monostatic scene classification. The overall system architecture is shown in Fig.~\ref{Fig1}. We first describe in Sec.~\ref{subsec_WaveDomain} the analog wave-domain mapping from a discretized scene reflectivity distribution to complex microwave measurements using an experimentally calibrated MNT model of the DMA. Next, we model in Sec.~\ref{subsec_ADC} the ADC stage that converts the complex analog measurements into a real-valued quantized feature vector. Then, we define in Sec.~\ref{subsec_DigitalDomain} the lightweight digital classifier. Finally, we summarize in Sec.~\ref{subsec_E2EPipeline} the resulting end-to-end sensing pipeline from scene reflectivity to predicted class label.

\subsection{Analog Wave Domain}
\label{subsec_WaveDomain}

The analog wave domain maps a scene reflectivity distribution to a sequence of complex microwave measurements through the DMA. We describe this mapping in three steps. \textit{First}, we introduce the MNT-based model of the DMA and
its binary tunable loads in Sec.~\ref{subsubsec_DMA}. \textit{Second}, we specify the discretized scene model used for the MNIST scene-classification task in Sec.~\ref{subsubsec_Scene}. \textit{Third}, we formalize our monostatic measurement model involving a single transmit feed and a single receive feed in Sec.~\ref{subsubsec_MonostaticMeasurementModel}.

\subsubsection{Dynamic Metasurface Antenna}
\label{subsubsec_DMA}

The MNT-based DMA model described next applies to all DMA architectures whose reconfigurability originates from tunable lumped elements. In Sec.~\ref{subsec_DMA_Setup}, we provide details of the specific chaotic-cavity-based DMA prototype considered in this work.

\begin{figure}
    \centering
    \includegraphics[width=0.8\columnwidth]{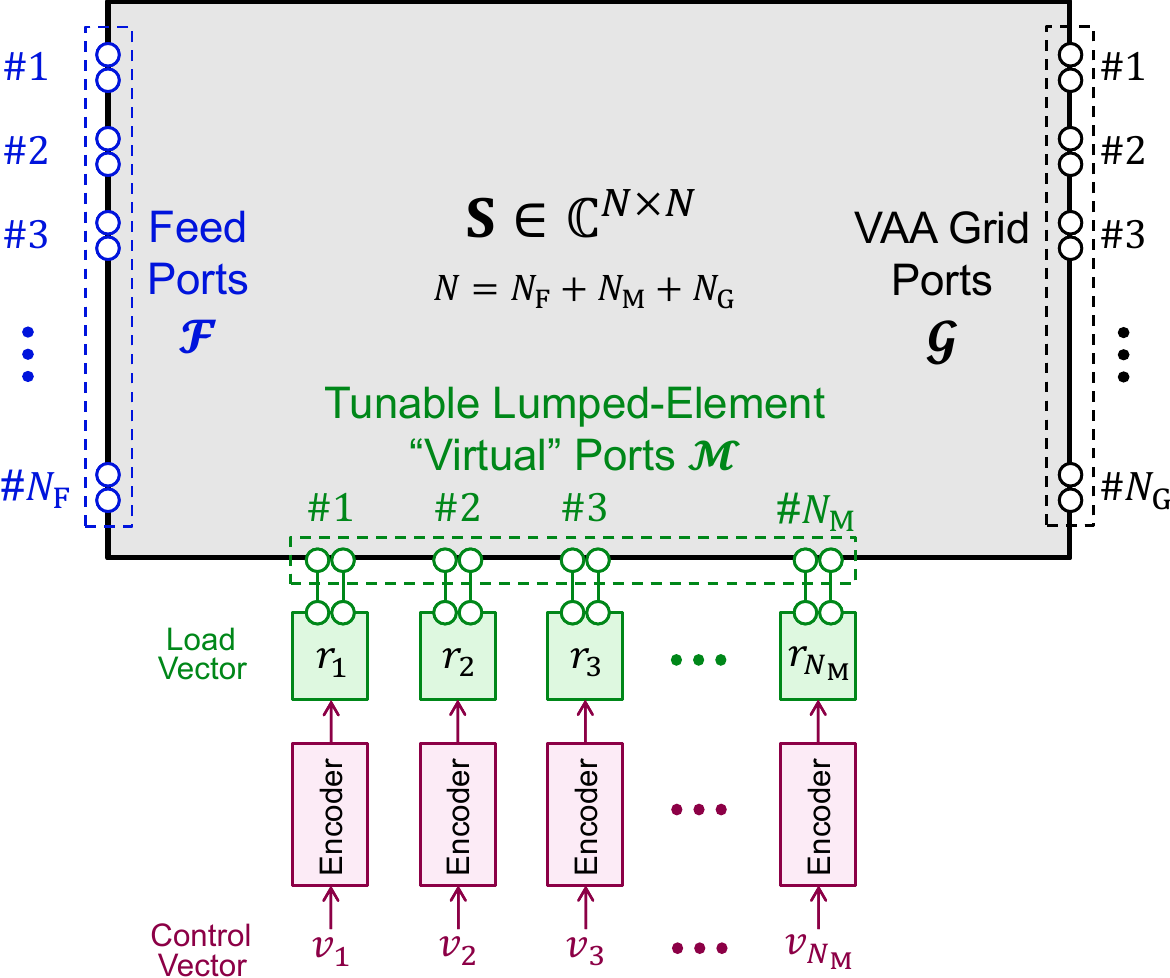}
\caption{Schematic illustration of the MNT-based model for a generic multi-feed DMA together with a VAA sampling the DMA's radiated field. Each port is depicted as a pair of two terminals.}
    \label{Fig2}
\end{figure}

Following~\cite{tapie2025experimental}, we model the DMA together with the virtual antenna array (VAA) used to sample its radiated field as a tunable multiport network. A schematic illustration is provided in Fig.~\ref{Fig2}. The VAA samples the DMA's radiated field at a finite set of spatial and polarization channels (see bottom panel of Fig.~\ref{Fig3}); in practice, the VAA is often implemented by mechanically displacing a probe. 
The DMA is a radiating structure with $N_\mathrm{F}$ feed ports and $N_\mathrm{M}$ tunable lumped elements; the VAA samples the DMA's radiation pattern at $N_\mathrm{G}$ grid points. Each of the DMA's tunable lumped elements can be represented as a ``virtual'' port terminated by an individual tunable load. The tunable multiport network comprising DMA and VAA can be partitioned into a static subsystem and a tunable subsystem. The static subsystem is characterized by its scattering matrix $\mathbf S \in \mathbb{C}^{N \times N}$, where $N=N_\mathrm{F}+N_\mathrm{M}+N_\mathrm{G}$. The tunable subsystem is the ensemble of the $N_\mathrm{M}$ tunable loads and characterized by its scattering matrix $\mathbf{\Phi} =\mathrm{diag}(\mathbf r) \in \mathbb{C}^{N_\mathrm{M}\times N_\mathrm{M}}$, where $\mathbf r=[r_1,\ldots,r_{N_\mathrm M}]^\top$ and $r_i\in\mathbb{C}$ is the reflection coefficient of the $i$th tunable load. The two subsystems are connected via the $N_\mathrm{M}$ ``virtual'' ports. Standard MNT yields the transmission matrix $\mathbf{T}\in\mathbb{C}^{N_\mathrm{G}\times N_\mathrm{F}}$ from DMA feed ports to the VAA as well as the reflection matrix $\mathbf{R}\in\mathbb{C}^{N_\mathrm{F}\times N_\mathrm{F}}$ at the DMA's feed ports:
\begin{subequations}
\label{eqMNT}
\begin{align}
\mathbf T(\mathbf r)
&=
\mathbf T_0
+
\mathbf A
\bigl(
\mathbf I_{N_\mathrm M}
-
\mathbf\Phi(\mathbf r)\mathbf\Gamma
\bigr)^{-1}
\mathbf\Phi(\mathbf r)
\mathbf B ,
\label{eq_DMA_MNT_T}
\\
\mathbf R(\mathbf r)
&=
\mathbf R_0
+
\mathbf B^\top
\bigl(
\mathbf I_{N_\mathrm M}
-
\mathbf\Phi(\mathbf r)\mathbf\Gamma
\bigr)^{-1}
\mathbf\Phi(\mathbf r)
\mathbf B ,
\label{eq_DMA_MNT_R}
\end{align}
\end{subequations}
where $\mathbf T_0\triangleq\mathbf S_{\mathcal G\mathcal F}\in\mathbb C^{N_\mathrm G\times N_\mathrm F}$, $\mathbf R_0\triangleq\mathbf S_{\mathcal F\mathcal F}\in\mathbb C^{N_\mathrm F\times N_\mathrm F}$, $\mathbf A\triangleq\mathbf S_{\mathcal G\mathcal M}\in\mathbb C^{N_\mathrm G\times N_\mathrm M}$, $\mathbf B\triangleq\mathbf S_{\mathcal M\mathcal F}\in\mathbb C^{N_\mathrm M\times N_\mathrm F}$, and $\mathbf\Gamma\triangleq\mathbf S_{\mathcal M\mathcal M}\in\mathbb C^{N_\mathrm M\times N_\mathrm M}$, and $\mathcal F$, $\mathcal M$, and $\mathcal G$ denote the port-index sets associated with the feeds, tunable elements, and VAA ports, respectively. 
We use $\mathbf S_{\mathcal F\mathcal M}=\mathbf S_{\mathcal M\mathcal F}^{\top}$ in \eqref{eqMNT} because the DMA and the free-space radio environment are reciprocal.
We assume throughout that $\mathbf I_{N_\mathrm M}-\mathbf\Phi(\mathbf r)\mathbf\Gamma$ is nonsingular for every admissible DMA configuration.
We emphasize that \eqref{eqMNT} is valid irrespective of the detailed DMA architecture (as long as the tunable elements are lumped) and the VAA details.\footnote{While we consider a single-polarization planar VAA in this paper (see bottom panel of Fig.~\ref{Fig3}), the same model can be applied to different VAA geometries and polarization details; for instance, \cite{tapie2026channel} applies the model to a spherical dual-polarized VAA (treating the two orthogonal polarizations as two separate VAA ports).} Moreover, we make no assumption about the separation between DMA and VAA; in particular, this means that \eqref{eqMNT} does \textit{not} require a far-field assumption. 

Many DMA prototypes, including ours, use binary tunable elements such as PIN diodes. We therefore parametrize the DMA configuration by a binary control vector $\mathbf v\in\mathbb B^{N_\mathrm M}$ and use the affine encoding
\begin{equation}
\mathbf r(\mathbf v)
=
\alpha\mathbf 1
+
(\beta-\alpha)\mathbf v ,
\label{eq_DMA_encoding}
\end{equation}
where $\alpha,\beta\in\mathbb C$ are the two load reflection coefficients associated with the two states of each tunable element. 

To assess the importance of MC awareness, we also consider an MC-unaware benchmark model obtained by setting $\mathbf\Gamma=\mathbf 0$ in \eqref{eqMNT}, namely
\begin{subequations}
\label{eqnoMC}
\begin{align}
\mathbf T^\mathrm{noMC}(\mathbf r)
&=
\mathbf T_0
+
\mathbf A
\mathbf\Phi(\mathbf r)
\mathbf B ,
\label{eq_DMA_noMC_T}
\\
\mathbf R^\mathrm{noMC}(\mathbf r)
&=
\mathbf R_0
+
\mathbf B^\top
\mathbf\Phi(\mathbf r)
\mathbf B .
\label{eq_DMA_noMC_R}
\end{align}
\end{subequations}

The parameters involved in the MNT-based DMA model are thus $\alpha$, $\beta$, $\mathbf T_0$, $\mathbf R_0$, $\mathbf A$, $\mathbf B$, and $\mathbf\Gamma$. As mentioned in our introduction, they are typically not known for an experimental DMA prototype. As discussed in~\cite{tapie2025experimental}, unambiguously estimating these parameters based on measurements of $\mathbf{T}$ and $\mathbf{R}$ for known $\mathbf{v}$ is not possible. Nonetheless, a set of proxy parameters can be estimated that accurately maps $\mathbf{v}$ to $\mathbf{T}$ and $\mathbf{R}$~\cite{tapie2025experimental}. Hence, the experimentally estimated proxy parameters are operationally equivalent to the ground-truth parameters for the purposes of the present work.

\subsubsection{Scene}
\label{subsubsec_Scene}

We model the scene as a discretized scalar reflectivity distribution $\boldsymbol\rho\in[0,1]^{N_\mathrm G}$ defined on the same grid as the VAA ports introduced above. The $g$th entry $\rho_g$ denotes the reflectivity associated with the $g$th VAA port. In line with the MNIST-based sensing task considered in this paper, $\boldsymbol\rho$ is a real-valued nonnegative grayscale reflectivity map. Thus, the scene model describes variations in scattering strength while using a common phase for all pixels.

Consistent with the first-Born approximation that is routinely used in computational DMA-based imaging~\cite{sleasman2015dynamic,sleasman2020implementation,del2020learned,qian2022noise}, we neglect multiple scattering between distinct scene pixels. Each scene pixel is therefore treated as an independent weak scatterer. 

For the MNIST classification task considered in this paper, we normalize each $28\times28$ MNIST image to $[0,1]$, center-crop it to $24\times24$, and downsample it to the $11\times11$ VAA grid by area averaging. The resulting $121$ pixel values are vectorized into $\boldsymbol\rho$. Hence, bright MNIST strokes correspond to more strongly reflecting scene pixels, whereas dark background pixels correspond to weakly reflecting scene pixels.

\subsubsection{Monostatic Measurement Model}
\label{subsubsec_MonostaticMeasurementModel}

Our shared-aperture monostatic measurements use a single DMA feed for transmission and a distinct single DMA feed for reception, implying the requirement $N_\mathrm{F}\geq 2$. If $N_\mathrm{F}>2$ (as for our prototype with $N_\mathrm{F}=7$), we partition the feed ports into two used feeds and the remaining unused feeds. We denote the corresponding port-index sets by $\mathcal A$ and $\mathcal U=\mathcal F\setminus\mathcal A$, respectively. We terminate the $N_\mathrm{F}-2$ unused feeds with open-circuit loads to avoid unnecessary signal loss through DMA feed ports that are not connected to RF chains. Since this termination is fixed and independent of the DMA configuration, it can be absorbed into an effective static multiport network with only the two 
feeds, the $N_\mathrm M$ tunable-element ports, and the $N_\mathrm G$ VAA ports.
Specifically, let
$\bar{\mathcal P}\triangleq\mathcal A\cup\mathcal M\cup\mathcal G$ denote the retained port set. Then, we obtain the scattering matrix $\bar{\mathbf S}\in \mathbb{C}^{N_\mathrm{P}\times N_\mathrm{P}}$ of the reduced static network, where $N_\mathrm{P}=2+N_\mathrm{M}+N_\mathrm{G}$, by eliminating the
unused feed ports $\mathcal U$ under open-circuit boundary conditions using standard MNT:
\begin{equation}
\bar{\mathbf S}
\triangleq
\mathbf S_{\bar{\mathcal P}\bar{\mathcal P}}
+
\mathbf S_{\bar{\mathcal P}\mathcal U}
\left(
\mathbf I_{|\mathcal U|}
-
\mathbf S_{\mathcal U\mathcal U}
\right)^{-1}
\mathbf S_{\mathcal U\bar{\mathcal P}}.
\label{eq_open_circuit_static_reduction}
\end{equation}
Consequently, the counterpart of \eqref{eqMNT} for this reduced static network is
\begin{subequations}
\label{eqMNT_reduced}
\begin{align}
\bar{\mathbf T}(\mathbf r)
&=
\bar{\mathbf T}_0
+
\bar{\mathbf A}
\bigl(
\mathbf I_{N_\mathrm M}
-
\mathbf\Phi(\mathbf r)\bar{\mathbf\Gamma}
\bigr)^{-1}
\mathbf\Phi(\mathbf r)
\bar{\mathbf B},
\label{eq_DMA_MNT_reduced_T}
\\
\bar{\mathbf R}(\mathbf r)
&=
\bar{\mathbf R}_0
+
\bar{\mathbf B}^{\top}
\bigl(
\mathbf I_{N_\mathrm M}
-
\mathbf\Phi(\mathbf r)\bar{\mathbf\Gamma}
\bigr)^{-1}
\mathbf\Phi(\mathbf r)
\bar{\mathbf B},
\label{eq_DMA_MNT_reduced_R}
\end{align}
\end{subequations}
where
$\bar{\mathbf T}(\mathbf r)\in\mathbb C^{N_\mathrm G\times 2}$ and
$\bar{\mathbf R}(\mathbf r)\in\mathbb C^{2\times2}$ are the feed-to-VAA
transmission matrix and feed-reflection matrix seen from the two used feeds, respectively.
Here,
$\bar{\mathbf T}_0\triangleq\bar{\mathbf S}_{\mathcal G\mathcal A}$,
$\bar{\mathbf R}_0\triangleq\bar{\mathbf S}_{\mathcal A\mathcal A}$,
$\bar{\mathbf A}\triangleq\bar{\mathbf S}_{\mathcal G\mathcal M}$,
$\bar{\mathbf B}\triangleq\bar{\mathbf S}_{\mathcal M\mathcal A}$, and
$\bar{\mathbf\Gamma}\triangleq\bar{\mathbf S}_{\mathcal M\mathcal M}$.

Next, we further partition the two used feeds into the transmit feed and the receive feed. We denote by $\mathcal{T}$ and $\mathcal{R}=\mathcal{A}\setminus\mathcal{T}$ the corresponding port indices. 
This allows us to extract the following three quantities of interest:
\begin{subequations}
\label{eq_monostatic_quantities}
\begin{align}
\mathbf t_{\mathcal G\mathcal T}(\mathbf r)
&\triangleq
\bar{\mathbf T}_{\mathcal G\mathcal T}(\mathbf r)
\in\mathbb C^{N_\mathrm G},
\label{eq_tGT}
\\
\mathbf t_{\mathcal G\mathcal R}(\mathbf r)
&\triangleq
\bar{\mathbf T}_{\mathcal G\mathcal R}(\mathbf r)
\in\mathbb C^{N_\mathrm G},
\label{eq_tGR}
\\
s_{\mathcal R\mathcal T}(\mathbf r)
&\triangleq
\bar{\mathbf R}_{\mathcal R\mathcal T}(\mathbf r)
\in\mathbb C .
\label{eq_sRT}
\end{align}
\end{subequations}
Here, $\mathbf t_{\mathcal G\mathcal T}(\mathbf r)$ is the field radiated on the VAA grid when exciting the transmit feed, whereas
$\mathbf t_{\mathcal G\mathcal R}(\mathbf r)$ is the corresponding VAA field associated with the receive feed. By reciprocity, the latter also characterizes the coupling from the VAA grid back to the receive feed. Finally, $s_{\mathcal R\mathcal T}(\mathbf r)$ is the direct TX-to-RX self-interference term through the DMA in the absence of the scene.

Under the routinely used first Born approximation~\cite{sleasman2015dynamic,sleasman2020implementation,del2020learned,qian2022noise}, which excludes multiple scattering within the scene and multiple scattering between the DMA and the scene, and assuming a unit-amplitude incident wave at the transmit feed for each DMA configuration, the received signal for the $m$th DMA configuration $\mathbf v_m$ is the superposition of three contributions: (i) the single-scene-bounce signal that propagates from the DMA's transmit port via the scene to the DMA's receive port, (ii) the self-interference signal that propagates from the DMA's transmit port to the DMA's receive port without interacting with the scene, and (iii) the additive measurement noise:
\begin{equation}
\tilde z_m(\boldsymbol\rho)
=
\boldsymbol\rho^\top
\left[
\mathbf t_{\mathcal G\mathcal R}(\mathbf r_m)
\odot
\mathbf t_{\mathcal G\mathcal T}(\mathbf r_m)
\right]
+
s_{\mathcal R\mathcal T}(\mathbf r_m)
+
n_m ,
\label{eq_single_measurement}
\end{equation}
where $\mathbf r_m\triangleq\mathbf r(\mathbf v_m)$ and $n_m\in\mathbb{C}$ denotes the additive measurement noise.

In the present work, we assume that the TX-to-RX self-interference term is ideally cancelled in the analog part of the receive chain, i.e., before the ADC stage. This assumption is natural in our setting because the self-interference term is independent of the scene reflectivity under the first-Born approximation and can be accurately predicted from the DMA's experimentally calibrated proxy MNT model, as seen in \eqref{eq_sRT}. In practice, such analog self-interference cancellation can be implemented by splitting off a reference copy of the transmitted signal, applying a configuration-dependent attenuation and phase shift, and injecting the resulting cancellation signal into the analog receive chain such that it destructively interferes with the direct TX-to-RX self-interference term before the ADC stage. Mathematically, this corresponds to subtracting the known complex offset $s_{\mathcal R\mathcal T}(\mathbf r_m)$ from the received signal for each DMA configuration. We assume ideal self-interference cancellation in order to focus on the interaction between the scene-dependent wave-domain measurements, finite-resolution ADC quantization, and downstream classification. Accordingly, after model-based self-interference cancellation, the effective analog measurement used by the ADC and digital classifier is
\begin{equation}
\begin{aligned}
z_m(\boldsymbol\rho)
&=
\tilde z_m(\boldsymbol\rho)-s_{\mathcal R\mathcal T}(\mathbf r_m)
\\
&=
\boldsymbol\rho^\top
\left[
\mathbf t_{\mathcal G\mathcal R}(\mathbf r_m)
\odot
\mathbf t_{\mathcal G\mathcal T}(\mathbf r_m)
\right]
+
n_m .
\end{aligned}
\label{eq_single_measurement_si_cancelled}
\end{equation}

For the MC-unaware benchmark with model-based self-interference cancellation, we compute the cancellation signal by first evaluating the full-feed MC-unaware model in \eqref{eqnoMC}, then imposing the same open-circuit termination on the unused feed ports and extracting the TX-to-RX entry; we denote the resulting prediction by $s^\mathrm{noMC}_{\mathcal R\mathcal T}(\mathbf r_m)$. During testing, we use the MC-aware physical forward model with the imperfect MC-unaware self-interference cancellation:
\begin{equation}
z_m^\mathrm{noMC}(\boldsymbol\rho) = \tilde z_m(\boldsymbol\rho)-s^\mathrm{noMC}_{\mathcal R\mathcal T}(\mathbf r_m).
\label{eq_single_measurement_si_cancelled_woMC}
\end{equation}

For a sequence of $M$ DMA configurations collected in
$\mathbf V=[\mathbf v_1,\ldots,\mathbf v_M]\in
\mathbb B^{N_\mathrm M\times M}$, the analog wave-domain measurement vector is
\begin{equation}
\mathbf z(\boldsymbol\rho,\mathbf V)
=
\left[
z_1(\boldsymbol\rho),\ldots,z_M(\boldsymbol\rho)
\right]^\top
\in\mathbb C^M .
\label{eq_complex_measurement_vector}
\end{equation}

\subsection{Analog-to-Digital Conversion}
\label{subsec_ADC}

Since the analog wave-domain measurements in \eqref{eq_complex_measurement_vector} are complex-valued, we assume coherent IQ reception and apply real-valued ADCs separately to the real and imaginary parts. Within each branch, the same scalar ADC is applied componentwise to all $M$ measurements. Before quantization, the analog front-end applies offset subtraction and gain control, separately to the I and Q branches, in order to align the measured signal distribution with the available ADC input range. We model these operations using branch-dependent calibration parameters for the ADC centers $c_{\chi}\in\mathbb R$ and dynamic ranges $\gamma_{\chi}>0$, where $\chi\in\{\mathrm I,\mathrm Q\}$. Specifically, $c_\chi$ and $\gamma_\chi$ are calibrated from the corresponding unnormalized analog branch samples over all training scenes and all measurement indices $m=1,\ldots,M$, namely from $\Re\{z_m\}$ for $\chi=\mathrm I$ and from $\Im\{z_m\}$ for $\chi=\mathrm Q$. We set $c_\chi$ to the corresponding empirical mean and $\gamma_\chi=2\eta\sigma_\chi$, where $\sigma_\chi$ is the corresponding empirical standard deviation and $\eta>0$ is a fixed range factor specified in Sec.~\ref{subsec_TrainingProcedure}.

We keep the dynamic ranges $\gamma_\chi$ fixed after calibration so that all ADC settings use the same input scale. For \textit{random-DMA} cases, we also keep the offsets $c_\chi$ fixed. For \textit{learned-DMA} cases, we recompute the offsets $c_\chi$ during training as deterministic calibration quantities obtained from the current front-end response; the offsets are thus not optimized. Specifically, for each current DMA front end, we set $c_\chi$ to the training-set mean of the corresponding unnormalized analog branch signal. Since the analog measurement model is linear in the scene reflectivity under the first-Born approximation and the additive noise is zero-mean, this training-set mean is the expected branch response for the current front end. After model selection, we freeze the selected offsets and dynamic ranges and use them without recalibration.

The analog offset subtraction and gain control map the complex wave-domain measurement $z_m$ to the normalized ADC inputs
\begin{subequations}
\label{eq_adc_normalized_input}
\begin{align}
u_{m,\mathrm I}
&=
2\left(\Re\{z_m\}-c_{\mathrm I}\right) / \gamma_{\mathrm I},
\label{eq_adc_normalized_input_I}
\\
u_{m,\mathrm Q}
&=
2\left(\Im\{z_m\}-c_{\mathrm Q}\right) / \gamma_{\mathrm Q},
\label{eq_adc_normalized_input_Q}
\end{align}
\end{subequations}
for $m=1,\ldots,M$. Thus, the unnormalized analog branch interval
$[c_\chi-\gamma_\chi/2,c_\chi+\gamma_\chi/2]$
is mapped to the normalized ADC input range $[-1,1]$. All ADC thresholds and reconstruction levels below are defined in this normalized ADC-input domain.

For an $L$-level ADC, where $L=2^B$ for $B$ bits, the bin boundaries are strictly ordered:
\begin{equation}
-1
=
\tau_{\chi,0}
<
\tau_{\chi,1}
<
\cdots
<
\tau_{\chi,L-1}
<
\tau_{\chi,L}
=
1 .
\label{eq_adc_threshold_ordering}
\end{equation}
We collect the bin boundaries of branch $\chi$ in
$\boldsymbol\tau_{\chi}
\triangleq
[
\tau_{\chi,0},
\tau_{\chi,1},
\ldots,
\tau_{\chi,L}
]^\top
\in\mathbb R^{L+1}$, and define the corresponding quantization intervals as
\begin{equation}
\mathcal I_{\chi,\ell}
=
\begin{cases}
[\tau_{\chi,0},\tau_{\chi,1}], & \ell=1,\\
(\tau_{\chi,\ell-1},\tau_{\chi,\ell}], & \ell=2,\ldots,L.
\end{cases}
\label{eq_adc_intervals}
\end{equation}
The reconstruction levels corresponding to the bins are
$q_{\chi,1},\ldots,q_{\chi,L}\in\mathbb R$, and we collect them in
$\mathbf q_{\chi}
\triangleq
[
q_{\chi,1},
\ldots,
q_{\chi,L}
]^\top
\in\mathbb R^L$.

The normalized ADC input is clipped,
\begin{equation}
\bar u_{m,\chi}
=
\min\left\{
\max\left\{
u_{m,\chi},
-1
\right\},
1
\right\},
\label{eq_adc_clipped_input}
\end{equation}
and quantized as
\begin{equation}
Q_{\boldsymbol\tau_{\chi},\mathbf q_{\chi}}
(u_{m,\chi})
=
q_{\chi,\ell}
\quad
\text{if }
\bar u_{m,\chi}\in\mathcal I_{\chi,\ell},
\qquad
\ell=1,\ldots,L .
\label{eq_scalar_adc_branch}
\end{equation}
For a vector input
$\mathbf u_{\chi}
=
[
u_{1,\chi},
\ldots,
u_{M,\chi}
]^\top
\in\mathbb R^M$, we define the corresponding componentwise quantizer as
\begin{equation}
\mathcal Q_{\boldsymbol\tau_{\chi},\mathbf q_{\chi}}
(\mathbf u_{\chi})
\triangleq
\left[
Q_{\boldsymbol\tau_{\chi},\mathbf q_{\chi}}(u_{1,\chi}),
\ \cdots, \
Q_{\boldsymbol\tau_{\chi},\mathbf q_{\chi}}(u_{M,\chi})
\right]^\top .
\label{eq_componentwise_adc_branch}
\end{equation}

We consider two ADC parameterizations:
\begin{enumerate}
    \item \textit{Uniform ADC:}
    This case corresponds to uniformly spaced reconstruction levels over the normalized ADC input range $[-1,1]$, with internal decision thresholds placed halfway between adjacent reconstruction levels. For each branch $\chi\in\{\mathrm I,\mathrm Q\}$, the reconstruction levels are
    \begin{equation}
    q_{\chi,\ell}
    =
    -1
    +
    2\frac{\ell-1}{L-1},
    \qquad
    \ell=1,\ldots,L .
    \label{eq_uniform_adc_levels}
    \end{equation}
    The internal bin boundaries are placed halfway between adjacent reconstruction levels,
    \begin{equation}
    \tau_{\chi,\ell}
    =
    \frac{
    q_{\chi,\ell}
    +
    q_{\chi,\ell+1}
    }{2},
    \qquad
    \ell=1,\ldots,L-1,
    \label{eq_uniform_adc_thresholds}
    \end{equation}
    together with the outer clipping boundaries
    $\tau_{\chi,0}=-1$ and $\tau_{\chi,L}=1$. Thus, the bin boundaries obey the ordering in \eqref{eq_adc_threshold_ordering}. In the \textit{uniform-ADC} case, the ADC thresholds and reconstruction levels are fixed.

    \item \textit{Learned ADC:}
    The reconstruction levels are kept fixed to the uniformly spaced values in \eqref{eq_uniform_adc_levels}. The internal bin boundaries
    $\tau_{\chi,1},\ldots,\tau_{\chi,L-1}$, however, are trainable, subject to the ordering constraint in \eqref{eq_adc_threshold_ordering}. Thus, the learned ADC adapts the quantization regions while keeping the scalar ADC output alphabet fixed. The trainable ADC parameters are
    \begin{equation}
    \boldsymbol\theta_{\mathrm{ADC}}^\mathrm{L}
    =
    \left\{
    \tau_{\chi,1},\ldots,\tau_{\chi,L-1}
    \right\}_{\chi\in\{\mathrm I,\mathrm Q\}} .
    \label{eq_learned_adc_parameters}
    \end{equation}
\end{enumerate}

In both cases, we collect the full list of ADC-stage parameters in $\boldsymbol\theta_{\mathrm{ADC}}
\triangleq
\left\{
c_\chi,\gamma_\chi,
\tau_{\chi,1},\ldots,\tau_{\chi,L-1},
q_{\chi,1},\ldots,q_{\chi,L}
\right\}_{\chi\in\{\mathrm I,\mathrm Q\}}$. In the \textit{uniform-ADC} case, no ADC thresholds or reconstruction levels are optimized. In the \textit{learned-ADC} case, only the internal thresholds
$\boldsymbol\theta_{\mathrm{ADC}}^\mathrm{L}\subset\boldsymbol\theta_{\mathrm{ADC}}$
are trainable. The calibration quantities $c_\chi$ and $\gamma_\chi$ follow the update-and-freeze protocol described above and are not optimized.

Altogether, the quantized measurement vector provided to the subsequent digital processing stage is
\begin{equation}
\mathbf y(\boldsymbol\rho,\mathbf V;\boldsymbol\theta_{\mathrm{ADC}})
=
\begin{bmatrix}
\mathcal Q_{\boldsymbol\tau_{\mathrm I},\mathbf q_{\mathrm I}}
\left(
\mathbf u_{\mathrm I}
\right)
\\
\mathcal Q_{\boldsymbol\tau_{\mathrm Q},\mathbf q_{\mathrm Q}}
\left(
\mathbf u_{\mathrm Q}
\right)
\end{bmatrix}
\in\mathbb R^{2M},
\label{eq_adc_output_vector}
\end{equation}
where
$\mathbf u_{\mathrm I}
=
[
u_{1,\mathrm I},
\ldots,
u_{M,\mathrm I}
]^\top$
and
$\mathbf u_{\mathrm Q}
=
[
u_{1,\mathrm Q},
\ldots,
u_{M,\mathrm Q}
]^\top$
are obtained from the complex measurement vector $\mathbf z(\boldsymbol\rho,\mathbf V)$ via \eqref{eq_adc_normalized_input}.
Because each complex DMA measurement contributes one I and one Q scalar ADC output, a scene acquired with $M$ DMA configurations and $B$ bits per real-valued ADC conversion has an acquisition bit budget of $2BM$ bits.
Reducing \(B\) lowers the number of acquired bits per feature, whereas reducing \(M\) shortens the digital feature vector; both reduce the acquisition bit budget and thus the amount of data that must be transferred, stored, and processed downstream.

For benchmarking purposes, we also consider the case of an \textit{ideal ADC} in which the quantization and clipping operations are bypassed and the digital processing stage receives the unquantized, offset-and-gain-controlled IQ feature vector
\begin{equation}
\mathbf y_{\mathrm{ideal}}(\boldsymbol\rho,\mathbf V)
=
\begin{bmatrix}
\mathbf u_{\mathrm I}
\\
\mathbf u_{\mathrm Q}
\end{bmatrix}
\in\mathbb R^{2M}.
\label{eq_ideal_adc_feature_vector}
\end{equation}

\subsection{Digital Domain}
\label{subsec_DigitalDomain}

The digital domain maps the ADC output vector
$\mathbf y(\boldsymbol\rho,\mathbf V;\boldsymbol\theta_{\mathrm{ADC}})\in\mathbb R^{2M}$
defined in \eqref{eq_adc_output_vector} to $\kappa$ class scores, where
$\kappa=10$ in our MNIST scene-classification scenario. The first operation in the digital-domain classifier is a fixed, non-trainable per-feature affine normalization layer:
\begin{equation}
\mathbf y_{\mathrm D}(\boldsymbol\rho,\mathbf V;\boldsymbol\theta_{\mathrm{ADC}})
=
\left(\mathbf y(\boldsymbol\rho,\mathbf V;\boldsymbol\theta_{\mathrm{ADC}})-\mathbf d_{\mathrm c}\right)\oslash \mathbf d_{\mathrm s}.
\label{eq_post_adc_normalization}
\end{equation}
Before gradient-based training, the calibration vectors
$\mathbf d_{\mathrm c},\mathbf d_{\mathrm s}\in\mathbb R^{2M}$ are estimated from the ADC-output features obtained from the training set using the initial wave-domain front end. Specifically, denoting by $\mathbf y_i^{(0)}$ the ADC-output vector associated with training sample $i$ during this calibration step, for $j=1,\ldots,2M$ we set
\begin{align}
[\mathbf d_{\mathrm c}]_j
&=
\frac{1}{|\mathcal D_{\mathrm{tr}}|}
\sum_{(\boldsymbol\rho_i,k_i)\in\mathcal D_{\mathrm{tr}}}
[\mathbf y_i^{(0)}]_j,
\\
[\mathbf d_{\mathrm s}]_j
&=
\max\left\{
\sqrt{
\operatorname{Var}_{(\boldsymbol\rho_i,k_i)\in\mathcal D_{\mathrm{tr}}}
\left([\mathbf y_i^{(0)}]_j\right)
},
\epsilon_{\mathrm D}
\right\},
\label{eq_post_adc_normalization_parameters}
\end{align}
with a small $\epsilon_{\mathrm D}>0$ for numerical stability. The vectors
$\mathbf d_{\mathrm c}$ and $\mathbf d_{\mathrm s}$ are then kept fixed throughout training, validation, and testing. Keeping this post-ADC normalization fixed is natural because the ADC reconstruction levels are fixed in both the \textit{uniform-ADC} and \textit{learned-ADC} cases. Following~\cite{del2020learned,qian2022noise,vol2025learning}, we then apply a fully connected artificial neural network (ANN) with one hidden layer to $\mathbf y_{\mathrm D}$. The ANN itself contains no batch-normalization layer.
We collect the ANN's trainable parameters in
\begin{equation}
\boldsymbol\theta_{\mathrm D}
=
\left\{
\mathbf W_1,\mathbf b_1,\mathbf W_2,\mathbf b_2
\right\},
\label{eq_digital_parameters}
\end{equation}
where
$\mathbf W_1\in\mathbb R^{N_\mathrm h\times 2M}$,
$\mathbf b_1\in\mathbb R^{N_\mathrm h}$,
$\mathbf W_2\in\mathbb R^{\kappa\times N_\mathrm h}$, and
$\mathbf b_2\in\mathbb R^{\kappa}$, and $N_\mathrm h$ denotes the number of
hidden units. The digital mapping is
\begin{align}
\mathbf a
&=
f_\mathrm{D}
\left(
\mathbf y(\boldsymbol\rho,\mathbf V;\boldsymbol\theta_{\mathrm{ADC}});
{\boldsymbol\theta_{\mathrm D}}
\right)
\\
&=
\mathbf W_2
\operatorname{ReLU}
\left(
\mathbf W_1
\mathbf y_{\mathrm D}(\boldsymbol\rho,\mathbf V;\boldsymbol\theta_{\mathrm{ADC}})
+
\mathbf b_1
\right)
+
\mathbf b_2 ,
\label{eq_digital_mapping}
\end{align}
where $\operatorname{ReLU}$ is applied componentwise and $\mathbf a=[a_0,\ldots,a_{\kappa-1}]^\top\in\mathbb R^{\kappa}$ denotes the vector of class logits. The predicted class is
\begin{equation}
\hat k
=
\arg\max_{k\in\{0,\ldots,\kappa-1\}}
a_k .
\label{eq_predicted_class}
\end{equation}

\subsection{End-to-End Sensing Pipeline}
\label{subsec_E2EPipeline}

Combining the analog wave-domain model from Sec.~\ref{subsec_WaveDomain}, the ADC model from Sec.~\ref{subsec_ADC}, and the digital ANN from Sec.~\ref{subsec_DigitalDomain}, our complete end-to-end sensing pipeline shown in Fig.~\ref{Fig1} can be summarized as follows:
\begin{equation}
\boldsymbol\rho
\;\xrightarrow{\ \mathbf V\ }\;
\mathbf z
\;\xrightarrow{\ \boldsymbol\theta_{\mathrm{ADC}}\ }\;
\mathbf y
\;\xrightarrow{\ \boldsymbol\theta_{\mathrm D}\ }\;
\hat k .
\label{eq_e2e_pipeline}
\end{equation}
The optimization problems discussed in Sec.~\ref{sec_OptAlg} are concerned with optimizing different subsets of $\{\mathbf V,\boldsymbol\theta_{\mathrm{ADC}},\boldsymbol\theta_{\mathrm D}\}$ to maximize the scene-classification accuracy of the end-to-end pipeline.

\section{Optimization Algorithms}
\label{sec_OptAlg}

This section describes our optimization of the end-to-end sensing pipeline defined in Sec.~\ref{subsec_E2EPipeline}. In Sec.~\ref{subsec_TrainingObjective}, we introduce the supervised training objective and the validation/test accuracy metric. In Sec.~\ref{subsec_OptCases}, we define the optimization cases used to isolate the roles of DMA learning, ADC awareness, and ADC learning. Since the DMA configurations and the ADC quantizer both involve nondifferentiable operations, in Sec.~\ref{subsec_DMADiffRelax} we describe our differentiable relaxation of the binary DMA configurations, and in Sec.~\ref{subsec_ADCDiffRelax} we describe our differentiable relaxation of ADC quantization. Finally, in Sec.~\ref{subsec_TrainingProcedure}, we specify the calibration, training, and model-selection procedure used in all numerical experiments.

\subsection{Training Objective}
\label{subsec_TrainingObjective}

We consider the task of single-label scene classification. 
To train the optimizable components of our end-to-end sensing pipeline using a supervised approach, we have access to a labelled classification dataset $\mathcal D = \left\{ (\boldsymbol\rho_i,k_i) \right\}_{i=1}^{{N}_\mathrm{data}}$, where $\boldsymbol\rho_i\in[0,1]^{N_\mathrm G}$ is the $i$th scene reflectivity vector and $k_i\in\{0,\ldots,\kappa-1\}$ is its class label. We split the available dataset into disjoint training, validation, and test sets denoted by $\mathcal D_{\mathrm{tr}}$, $\mathcal D_{\mathrm{val}}$, and
$\mathcal D_{\mathrm{test}}$, respectively, as detailed in Sec.~\ref{subsec_TrainingProcedure}.

As described in Sec.~\ref{sec_SystemModel}, for a scene $\boldsymbol\rho_i$, the
end-to-end pipeline produces the logit vector $\mathbf a_i
=
f_\mathrm{D}\!\left(
\mathbf y(\boldsymbol\rho_i,\mathbf V;\boldsymbol\theta_{\mathrm{ADC}});
\boldsymbol\theta_{\mathrm D}
\right)$, where $a_{i,k}$ denotes the raw class score associated with class $k$. The corresponding softmax probability assigned to class $k$ is $p_{i,k}
= \exp(a_{i,k}) / \sum_{j=0}^{\kappa-1}\exp(a_{i,j})$.
Since our task is single-label classification, the desired behavior is that the
probability assigned to the correct class $k_i$ is large. We therefore train the
optimizable parameters by minimizing the average cross-entropy loss over the
training set, defined as
\begin{equation}
\mathcal J_{\mathrm{tr}}
\left(
\mathbf V,\boldsymbol\theta_{\mathrm{ADC}},\boldsymbol\theta_{\mathrm D}
\right)
=
-\frac{1}{|\mathcal D_{\mathrm{tr}}|}
\sum_{(\boldsymbol\rho_i,k_i)\in\mathcal D_{\mathrm{tr}}}
\log p_{i,k_i}.
\label{eq_training_objective}
\end{equation}

For validation and testing, we evaluate the classification accuracy $
\left|
\left\{
(\boldsymbol\rho_i,k_i)\in\mathcal D'
:
\hat k_i=k_i
\right\}
\right|
\big/
|\mathcal D'|$, where $\mathcal D'$ denotes either $\mathcal D_{\mathrm{val}}$ or $\mathcal D_{\mathrm{test}}$.

\begin{table*}[h]
\centering
\caption{Optimization cases used to isolate the roles of DMA learning, ADC awareness, and ADC-threshold learning. $\mathbf V_{\mathrm{rnd}}$ denotes a fixed randomly chosen $\mathbf{V}$.}\label{tab_opt_cases}
\renewcommand{\arraystretch}{1.45}
\setlength{\tabcolsep}{4pt}
\footnotesize
\begin{tabular}{@{}>{\raggedright\arraybackslash}p{0.15\textwidth}
>{\centering\arraybackslash}p{0.27\textwidth}
>{\centering\arraybackslash}p{0.27\textwidth}
>{\centering\arraybackslash}p{0.27\textwidth}@{}}
\hline
& \textit{Ideal ADC} & \textit{Uniform ADC} & \textit{Learned ADC} \\
\hline
\textit{Random DMA}
&
$\displaystyle
\min_{\boldsymbol\theta_{\mathrm D}}
\mathcal J_{\mathrm{tr}}^{\mathrm{ideal}}
(\mathbf V_{\mathrm{rnd}},\boldsymbol\theta_{\mathrm D})
$
&
$\displaystyle
\min_{\boldsymbol\theta_{\mathrm D}}
\mathcal J_{\mathrm{tr}}
(\mathbf V_{\mathrm{rnd}},\boldsymbol\theta_{\mathrm{ADC}},\boldsymbol\theta_{\mathrm D})
$
&
$\displaystyle
\min_{\boldsymbol\theta_{\mathrm{ADC}}^{\mathrm L},\boldsymbol\theta_{\mathrm D}}
\mathcal J_{\mathrm{tr}}
(\mathbf V_{\mathrm{rnd}},\boldsymbol\theta_{\mathrm{ADC}},\boldsymbol\theta_{\mathrm D})
$
\\[2.0ex]

\textit{Learned DMA}
&
$\displaystyle
\min_{\mathbf V,\boldsymbol\theta_{\mathrm D}}
\mathcal J_{\mathrm{tr}}^{\mathrm{ideal}}
(\mathbf V,\boldsymbol\theta_{\mathrm D})
$
&
$\displaystyle
\min_{\mathbf V,\boldsymbol\theta_{\mathrm D}}
\mathcal J_{\mathrm{tr}}
(\mathbf V,\boldsymbol\theta_{\mathrm{ADC}},\boldsymbol\theta_{\mathrm D})
$
&
$\displaystyle
\min_{\mathbf V,\boldsymbol\theta_{\mathrm{ADC}}^{\mathrm L},\boldsymbol\theta_{\mathrm D}}
\mathcal J_{\mathrm{tr}}
(\mathbf V,\boldsymbol\theta_{\mathrm{ADC}},\boldsymbol\theta_{\mathrm D})
$
\\
\hline
\end{tabular}
\end{table*}

\subsection{Optimization Cases}
\label{subsec_OptCases}

The different optimization cases detailed in Table~\ref{tab_opt_cases}
seek to minimize the training objective in \eqref{eq_training_objective}
with respect to different subsets of the three parameter groups
$\{\mathbf V,\boldsymbol\theta_{\mathrm{ADC}},\boldsymbol\theta_{\mathrm D}\}$.
For notational clarity, we denote by
$\mathcal J_{\mathrm{tr}}^{\mathrm{ideal}}(\mathbf V,\boldsymbol\theta_{\mathrm D})$
the training objective evaluated with
$\mathbf y=\mathbf y_{\mathrm{ideal}}(\boldsymbol\rho,\mathbf V)$ from
\eqref{eq_ideal_adc_feature_vector} instead of the quantized feature vector in
\eqref{eq_adc_output_vector}.

We further examine the importance of ADC awareness and MC awareness:

\textit{ADC awareness:} For the two cases in Table~\ref{tab_opt_cases} involving ideal ADCs, we additionally evaluate the optimized variables after replacing the ideal ADC by a finite-resolution uniform ADC. This evaluation quantifies the degradation caused by ignoring ADC quantization during optimization.
    
\textit{MC awareness:} For all cases in Table~\ref{tab_opt_cases}, we additionally perform model-based self-interference cancellation and training using the MC-unaware model in \eqref{eqnoMC}, and evaluate the resulting optimized variables using the MC-aware model in \eqref{eqMNT} while retaining the MC-unaware self-interference cancellation signal. This quantifies the degradation caused by ignoring MC during self-interference prediction and optimization.

\subsection{Differentiable Relaxation of Binary DMA Configurations}
\label{subsec_DMADiffRelax}

The learned-DMA optimization cases in Sec.~\ref{subsec_OptCases} require optimizing the binary DMA configuration matrix $\mathbf V\in\mathbb B^{N_\mathrm M\times M}$. Direct gradient-based optimization over this discrete set is not possible. We therefore use a soft-to-hard relaxation similar to the one used in~\cite{del2020learned,qian2022noise}, which in turn built on~\cite{chakrabarti2016learning}. 

\begin{figure}
    \centering
    \includegraphics[width=\columnwidth]{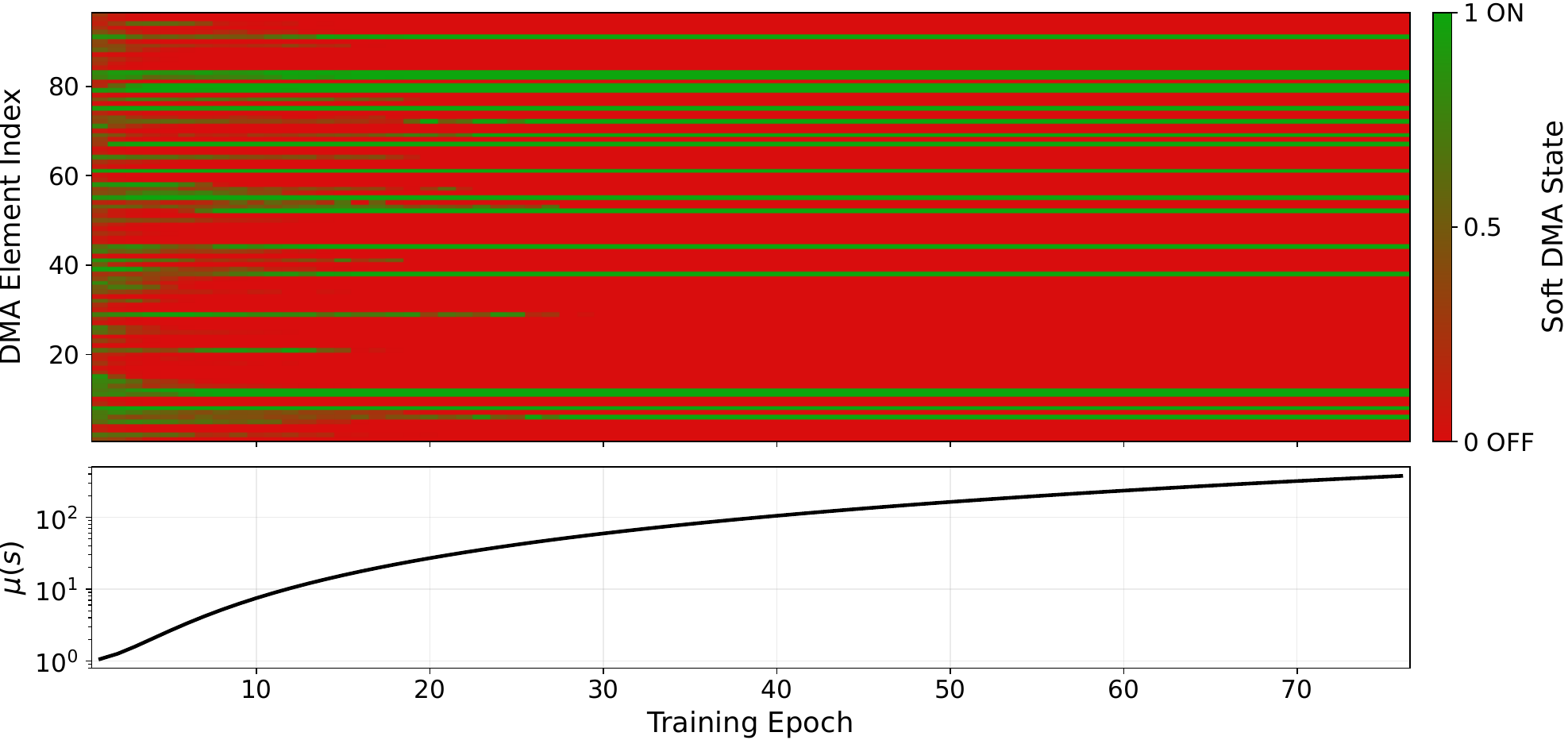}
    \caption{Soft-to-hard DMA relaxation for a representative \textit{learned-DMA learned-ADC} run with $M=1$ and $B=1$. Top panel: relaxed states $\widetilde v_{n,m}$ over training epochs. Bottom panel: hardening factor $\mu(s)$ evaluated at the end of each epoch. }
    \label{FigDMAbin}
\end{figure}

For each tunable element $n$ and measurement $m$, we introduce two trainable real-valued parameters $\omega_{n,m,0}$ and $\omega_{n,m,1}$ associated with the two hardware states. 
At training step $s$, we define the relaxed state of the \(n\)th tunable element during the \(m\)th measurement as
\begin{equation}
\widetilde v_{n,m}
=
\frac{
\exp\left(\mu(s)|\omega_{n,m,1}|\right)
}{
\exp\left(\mu(s)|\omega_{n,m,0}|\right)
+
\exp\left(\mu(s)|\omega_{n,m,1}|\right)
},
\label{eq_soft_dma_config}
\end{equation}
where $\mu(s)=1+(\gamma_\mathrm{DMA}s)^{p_\mathrm{DMA}}$ is a gradually increasing scale factor, equivalently an ``inverse temperature'', whose hyperparameters are specified in Sec.~\ref{subsec_TrainingProcedure}. For finite $\mu(s)$, $\widetilde v_{n,m}$ is differentiable almost everywhere with respect to the trainable parameters and can take intermediate values between $0$ and $1$.\footnote{Because of the absolute-value parametrization, nondifferentiabilities can occur only when $\omega_{n,m,0}=0$ or $\omega_{n,m,1}=0$. The isolated kinks do not require special treatment in practice, since automatic differentiation uses the standard subgradient of the absolute-value operation.} For validation and testing, we use the hardware-valid binary configuration $\check v_{n,m}$ obtained by setting $\check v_{n,m}=1$ if $|\omega_{n,m,1}|>|\omega_{n,m,0}|$ and $\check v_{n,m}=0$ otherwise.

During joint training, we evaluate the analog wave-domain with the MNT-based DMA model using the relaxed differentiable configuration $\widetilde{\mathbf V}$, i.e., the load vector of the $m$th measurement is computed from $\widetilde{\mathbf v}_m=[\widetilde v_{1,m},\ldots,\widetilde v_{N_\mathrm M,m}]^\top$ via \eqref{eq_DMA_encoding}. A representative evolution of the relaxed element states and of the hardening factor during training is shown in Fig.~\ref{FigDMAbin}.
For validation and testing (see further details in Sec.~\ref{subsec_TrainingProcedure}), we use the corresponding hard configuration $\check{\mathbf V}$.

\subsection{Differentiable Relaxation of ADC Quantization}
\label{subsec_ADCDiffRelax}

The ADC stage defined in Sec.~\ref{subsec_ADC} is also nondifferentiable, but the challenge differs from that associated with the DMA configuration in Sec.~\ref{subsec_DMADiffRelax}. For the DMA, nondifferentiability arises because the hardware configuration variables themselves are discrete. For the ADC, nondifferentiability arises from the hard bin assignment that maps each continuous input-signal sample to one of finitely many reconstruction levels. This staircase mapping obstructs backpropagation through the ADC stage, both when optimizing upstream variables such as the DMA configuration and when optimizing ADC parameters themselves. Following~\cite{shlezinger2022deep}, we therefore replace hard step functions by smooth hyperbolic-tangent transitions during training. The resulting surrogate is differentiable almost everywhere; at clipping boundaries, automatic differentiation provides a subgradient.

\begin{figure}
    \centering
    \includegraphics[width=\columnwidth]{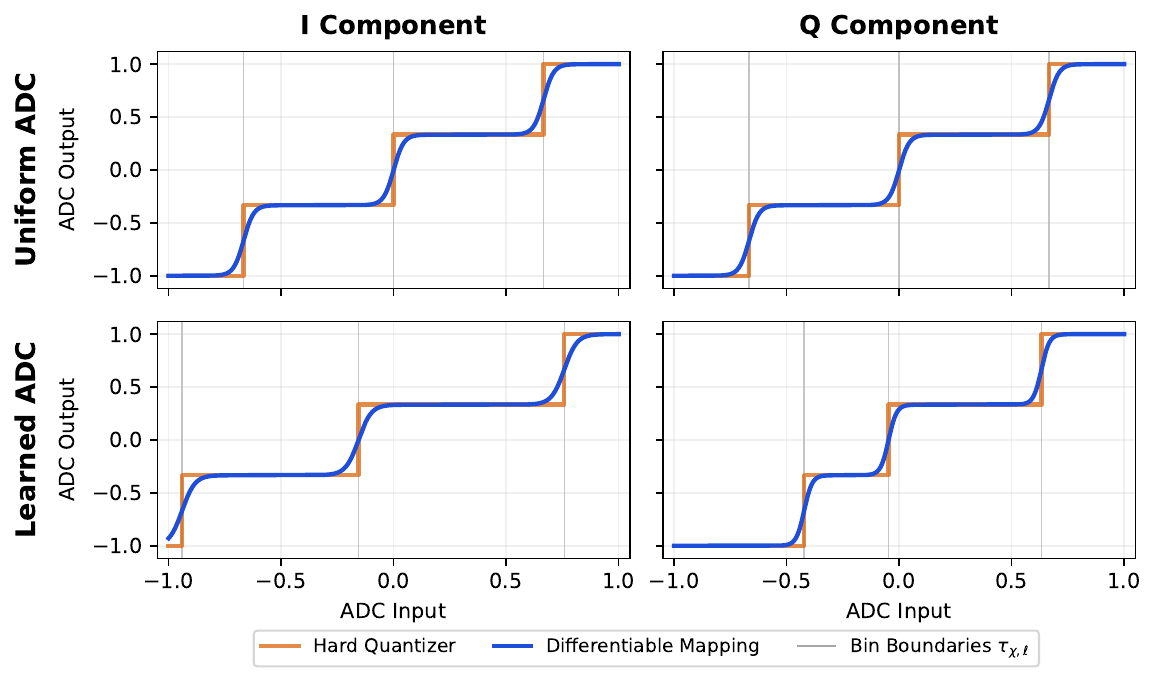}
\caption{Illustration of hard ADC quantization and its differentiable relaxation for $B=2$ in the \textit{learned-DMA} case with $M=1$. The orange staircases show the deployed hard quantizers $Q_{\boldsymbol\tau_\chi,\mathbf q_\chi}$, while the blue curves show the differentiable mappings $\widetilde Q_\chi$ used during training. The gray vertical lines indicate the bin boundaries $\tau_{\chi,\ell}$. Rows compare \textit{uniform-ADC} and \textit{learned-ADC} cases; columns show the corresponding I and Q branches.}
    \label{FigADCsoft}
\end{figure}

During training, we replace the ADC's hard bin transitions in \eqref{eq_scalar_adc_branch} by the differentiable surrogate
\begin{equation}
\widetilde Q_\chi(u)
=
q_{\chi,1}
+
\sum_{\ell=1}^{L-1}
(q_{\chi,\ell+1}-q_{\chi,\ell})
\frac{
1+\tanh\!\left(\lambda_{\mathrm{ADC}}\left(\bar u-\tau_{\chi,\ell}\right)\right)
}{2},
\label{eq_adc_soft_step_quantizer}
\end{equation}
where $\bar u=\min\{\max\{u,-1\},1\}$ is the clipped normalized ADC input, in agreement with \eqref{eq_adc_clipped_input}. The fixed slope $\lambda_{\mathrm{ADC}}>0$ controls the sharpness of the smooth transitions in the normalized ADC-input domain and is \textit{not} annealed during training; its numerical value is specified in Sec.~\ref{subsec_TrainingProcedure}. Larger values of $\lambda_{\mathrm{ADC}}>0$ make the surrogate closer to the corresponding hard quantizer.

In the \textit{uniform-ADC} baseline, both the reconstruction levels and the bin boundaries are fixed.
In the \textit{learned-ADC} case, the reconstruction levels remain fixed, while the internal bin boundaries are trained. Hence, the \textit{learned-ADC} case adapts the quantization regions to the downstream classification task. We keep the reconstruction levels fixed in order to retain a standard ADC output alphabet and isolate the effect of learning the decision thresholds.

To guarantee ordered bin boundaries throughout training in the \textit{learned-ADC} case, we parameterize the spacings between consecutive boundaries rather than the boundaries themselves. For each branch $\chi\in\{\mathrm I,\mathrm Q\}$, let $\xi_{\chi,1},\ldots,\xi_{\chi,L}$ be unconstrained trainable real-valued variables and define positive spacings $\Delta_{\chi,\ell}=\log(1+\exp(\xi_{\chi,\ell}))+\epsilon_\tau$, with a small $\epsilon_\tau>0$. Starting from the left clipping boundary, the internal bin boundaries are obtained cumulatively:
\begin{equation}
\tau_{\chi,\ell}
=
-1
+
2
\frac{\sum_{r=1}^{\ell}\Delta_{\chi,r}}{\sum_{r=1}^{L}\Delta_{\chi,r}},
\qquad
\ell=1,\ldots,L-1 .
\label{eq_learned_adc_ordered_thresholds}
\end{equation}
This construction guarantees
$-1<\tau_{\chi,1}<\cdots<\tau_{\chi,L-1}<1$, without sorting, projection, or auxiliary ordering penalties.

During joint training, we propagate gradients through the smooth quantizer in \eqref{eq_adc_soft_step_quantizer}. For validation and testing, we use the corresponding hard quantizer in \eqref{eq_scalar_adc_branch}, with the learned bin boundaries in the \textit{learned-ADC} case. In Fig.~\ref{FigADCsoft}, we illustrate a representative example of the hard quantizer used for validation and testing together with the differentiable surrogate used during training. The \textit{learned-ADC} case generally adapts the bin boundaries differently for the I and Q branches, reflecting their generally different signal distributions.

\subsection{Calibration, Training, and Model Selection}
\label{subsec_TrainingProcedure}

We use a standard MNIST split with $51000$ training samples, $9000$ validation samples, and $10000$ test samples. The training set is used to optimize the trainable parameters and to calibrate the fixed normalization parameters described below. The validation set is used only for model selection, and the test set is used only for the final performance evaluation.

Before training, we calibrate the pre-ADC affine transformation described in Sec.~\ref{subsec_ADC} from the training-set analog measurements. In all cases, we use $\eta=3$. The scale parameters $\gamma_\chi$ are kept fixed, while the offsets $c_\chi$ are fixed for \textit{random-DMA} cases and updated-then-frozen for \textit{learned-DMA} cases, as described in Sec.~\ref{subsec_ADC}. We also calibrate the post-ADC normalization in \eqref{eq_post_adc_normalization} from the training-set features obtained with the initial front end and the initial ADC parameters. This yields the per-feature vectors $\mathbf d_{\mathrm c}$ and $\mathbf d_{\mathrm s}$ in \eqref{eq_post_adc_normalization_parameters}, for which we use $\epsilon_{\mathrm D}=10^{-12}$. We keep these vectors fixed during training, validation, and testing. The ANN has $N_\mathrm h=256$ hidden units and contains no batch-normalization layer or other adaptive normalization.

For \textit{learned-DMA} cases, we optimize the relaxed DMA variables described in Sec.~\ref{subsec_DMADiffRelax}. The hyperparameters for the inverse-temperature parameter are $\gamma_{\mathrm{DMA}}=5\cdot10^{-4}$ and $p_{\mathrm{DMA}}=2$.

For the finite-resolution ADC cases (\textit{uniform-ADC} and \textit{learned-ADC}), we use the smooth ADC surrogate in \eqref{eq_adc_soft_step_quantizer} during gradient-based training. We keep the ADC slope parameter fixed and set it to $\lambda_{\mathrm{ADC}}=15/\Delta_{\mathrm{ADC}}$, where $\Delta_{\mathrm{ADC}}$ is a reference transition width in the normalized ADC-input domain. 
In the \textit{uniform-ADC} case, the ADC thresholds and reconstruction levels are fixed, and we use $\Delta_{\mathrm{ADC}}=2/L$.
In the \textit{learned-ADC} case, we initialize the internal bin boundaries from the corresponding uniform ADC and then optimize them jointly with the other trainable parameters; for the spacing floor in \eqref{eq_learned_adc_ordered_thresholds}, we use $\epsilon_\tau=10^{-6}$.
For this case, we use $\Delta_{\mathrm{ADC}}=2/(L-1)$ for $L>2$, corresponding to the initial spacing between adjacent reconstruction levels; for $L=2$, we use $\Delta_{\mathrm{ADC}}=2/L$. We use the Adam optimizer with learning rates $10^{-3}$ for the ANN, $2\cdot10^{-2}$ for the DMA parameters when applicable, and $10^{-3}$ for the learned ADC parameters when applicable.

During each epoch, we use mini-batches of $100$ training samples to update the trainable parameters. After each epoch, we evaluate the deployed hard system on the validation set, namely the hard DMA configuration and, for finite-resolution ADCs, the hard quantizer, using the corresponding hard-front-end calibration. We train for at most $240$ epochs and retain the parameter values and calibration quantities that yield the lowest validation cross-entropy loss after a minimum of $120$ training epochs. We stop training early if this validation loss does not improve for $15$ consecutive epochs after the minimum number of epochs. This hard-validation selection criterion avoids selecting a model that performs well only through the differentiable training relaxations.

We repeat each optimization case for ten independent random seeds. The seed controls the initialization of the DMA relaxation in \textit{learned-DMA} cases or the random DMA configurations in \textit{random-DMA} cases, the ANN weights, and the mini-batch order. For each seed, we first perform model selection independently using the validation set as described above. For each optimization case and parameter setting, we then report the test accuracy of the seed yielding the best validation performance. This common validation-based selection protocol accounts for run-to-run variability without using the test set for model selection. 

Finally, we evaluate the selected hard system once on the test set, using the selected hard DMA configurations, the hard ADC quantizer, the frozen pre-ADC calibration, the frozen post-ADC normalization, and the trained ANN without any recalibration on validation or test data.

\section{Results}
\label{sec_ExpResults}

This section presents our results for the proposed ADC-aware and MC-aware DMA-based end-to-end sensing framework. In Sec.~\ref{subsec_DMA_Setup}, we describe the fabricated chaotic-cavity-backed DMA prototype and its proxy MNT-based characterization, which provides the physical and modeling basis for all evaluations. In Sec.~\ref{subsec_SceneClassifResults}, we report scene-classification results across different operating regimes: ADC resolution constraints, single-shot operation ($M=1$), and pre-ADC noise.

\subsection{DMA Prototype and Experimental Setup}
\label{subsec_DMA_Setup}

\begin{figure}
    \centering
    \includegraphics[width=\columnwidth]{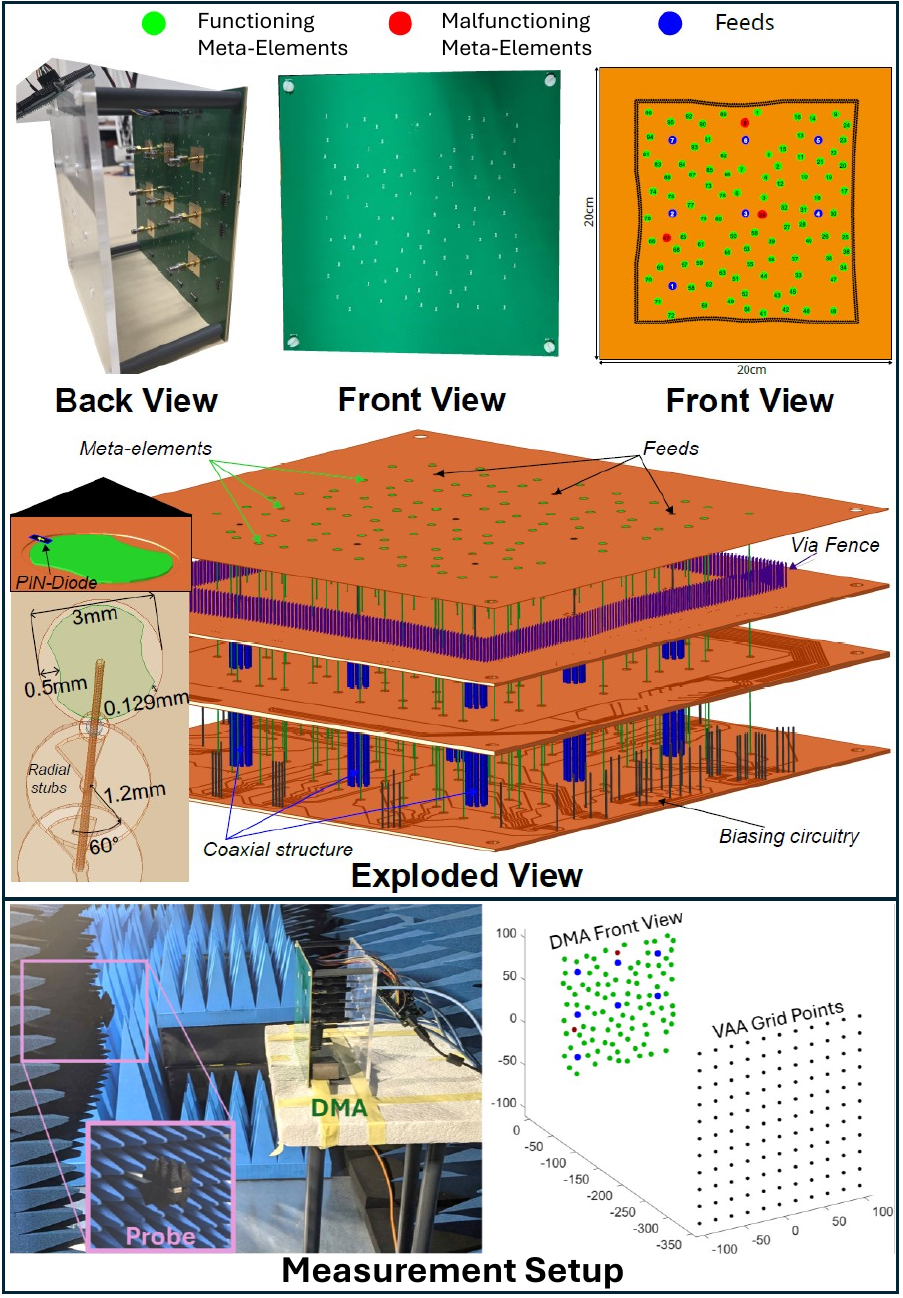}
\caption{DMA prototype and VAA-based experimental characterization setup. The top panel shows various views of the chaotic-cavity-backed multi-feed DMA prototype. The bottom panel shows the measurement setup that samples the DMA's radiated field at 19~GHz on an $11\times 11$ VAA grid located $35\,\mathrm{cm}$ in front of the DMA's aperture. Further details can be found in~\cite{tapie2025experimental}.}
    \label{Fig3}
\end{figure}

Our DMA prototype is based on the chaotic-cavity-backed architecture originally proposed in~\cite{sleasman2020implementation} because the resulting strong inter-element coupling boosts the DMA's wave-domain flexibility~\cite{prod2025mutual,prod2025benefits}. While the DMA prototype in~\cite{sleasman2020implementation} only had a single feed, ours has eight feeds to accommodate separate transmit and receive feeds, as well as auxiliary feeds used for the proxy MNT parameter estimation~\cite{tapie2025experimental}. As seen in the top panel in Fig.~\ref{Fig3}, a chaotic cavity couples the eight feeds to $N_\mathrm{M}=96$ one-bit programmable meta-elements. The quasi-two-dimensional cavity is formed by two parallel metallic layers and an irregular via fence, whose shape is chosen to break symmetries and promote wave-chaotic coupling among the feeds and meta-elements. Each feed excites the cavity from the back side through a coaxial connector. The meta-elements are patterned on the front side of the cavity and follow the complementary electric-LC (cELC) resonator design of~\cite{yoo2016efficient}. Each cELC resonator is loaded by a PIN diode and can be switched independently between two bias states. Bias voltages are delivered through vias and controlled electronically. We operate the DMA at 19~GHz. 

As displayed in the bottom panel in Fig.~\ref{Fig3}, we sample the DMA's radiated field using a VAA realized by mechanically scanning an open-ended waveguide probe in a plane parallel to the DMA aperture. The probe is oriented to receive the dominant polarization radiated by the DMA meta-elements. The VAA plane is located $35\,\mathrm{cm}$ in front of the DMA, and the scanned grid is an $11\times 11$ square grid with spacing $2\,\mathrm{cm}$, yielding $N_\mathrm{G}=121$ sampled field points. We recall that the MNT model does not require any specific choice of DMA-VAA separation or VAA geometry. Our goal here is not to infer a complete far-field radiation pattern or a modal expansion of the DMA aperture field, but to obtain an accurate finite-dimensional input-output model for this fixed VAA geometry. 

Our estimation of the proxy MNT parameters is based on measurements conducted with an eight-port vector network analyzer (VNA). One VNA port is connected to the VAA probe and the remaining seven VNA ports are connected to seven DMA feeds. The eighth DMA feed remains open-circuited throughout all experiments: it therefore plays the role of a static scatterer within the complex scattering environment of the chaotic cavity. As described in~\cite{tapie2025experimental}, we estimate a set of proxy MNT parameters from measurements of the reflected field at the seven used feeds and the radiated field on the VAA. Our characterization uses $1000$ random DMA configurations for the feed-reflection measurements and $60$ random DMA configurations for the VAA measurements, in addition to one reference configuration and $N_\mathrm{M}$ single-toggle configurations required by our segmented estimation procedure. 
We quantify the accuracy of the resulting calibrated proxy MNT model by its ability to predict the feed-reflection matrix and the feed-to-VAA transmission matrix for a sequence of 30 random and previously unseen DMA configurations. Our metric is defined like a signal-to-noise ratio (SNR) treating the prediction error as noise, as defined in [(18),~\cite{tapie2025experimental}], and reaches  $\zeta_\mathrm{R}=40.3\,\mathrm{dB}$ for the feed-reflection matrix $\mathbf R$ and $\zeta_\mathrm{T}=37.7\,\mathrm{dB}$ for the feed-to-VAA transmission matrix $\mathbf T$. These accuracies are remarkable given that, for $N_\mathrm{F}=7$, $N_\mathrm{M}=96$, and $N_\mathrm{G}=121$, the proxy MNT model contains $17{,}821$ complex-valued unknowns that are estimated only indirectly from measurements at accessible ports. Further details on the full calibration procedure can be found in~\cite{tapie2025experimental}.

\subsection{Scene Classification Results}
\label{subsec_SceneClassifResults}

We now present selected scene-classification results obtained with the proposed setup, together with a more detailed analysis of selected internal properties of the sensing pipeline. Our emphasis is not on reporting high absolute classification accuracies, but rather on qualitatively observing how the accuracy depends on key system parameters, in particular the number of measurements $M$ and the ADC resolution $B$. The most informative regimes are those in which the information flow through the pipeline is constrained, especially when the acquisition bit budget is small. For example, $M=1$ creates a bottleneck in the analog measurement stage because the scene must be classified from a single complex measurement, whereas $B=1$ creates a bottleneck at the ADC stage because each real-valued branch is reduced to a single binary decision. By contrast, when both $M$ and $B$ are large and the noise is not too strong, enough information reaches the ANN even for \textit{random-DMA uniform-ADC} scenarios, so that the marginal benefit of end-to-end learning becomes smaller and the resulting trends are less insightful. We therefore first examine the strongly ADC-limited regime with one-bit ADCs in Sec.~\ref{subsubsec_Exp1bitClassifResults}, and then the strongly measurement-limited regime with a single DMA configuration in Sec.~\ref{subsubsec_ExpSingleShotMeasurement}. In both cases, we assume negligible measurement noise, as in~\cite{del2020learned}, in order to isolate the effects of the analog and ADC bottlenecks. Since the measurement noise in the considered microwave measurements is additive Gaussian noise, its main expected effect is a systematic degradation of the classification accuracy~\cite{qian2022noise}; while noise-strength ablations are therefore not the central focus of the present section, we include an example thereof in Sec.~\ref{subsubsec_ablation_Noise}.

\begin{figure*}
    \centering
    \includegraphics[width=1.8\columnwidth]{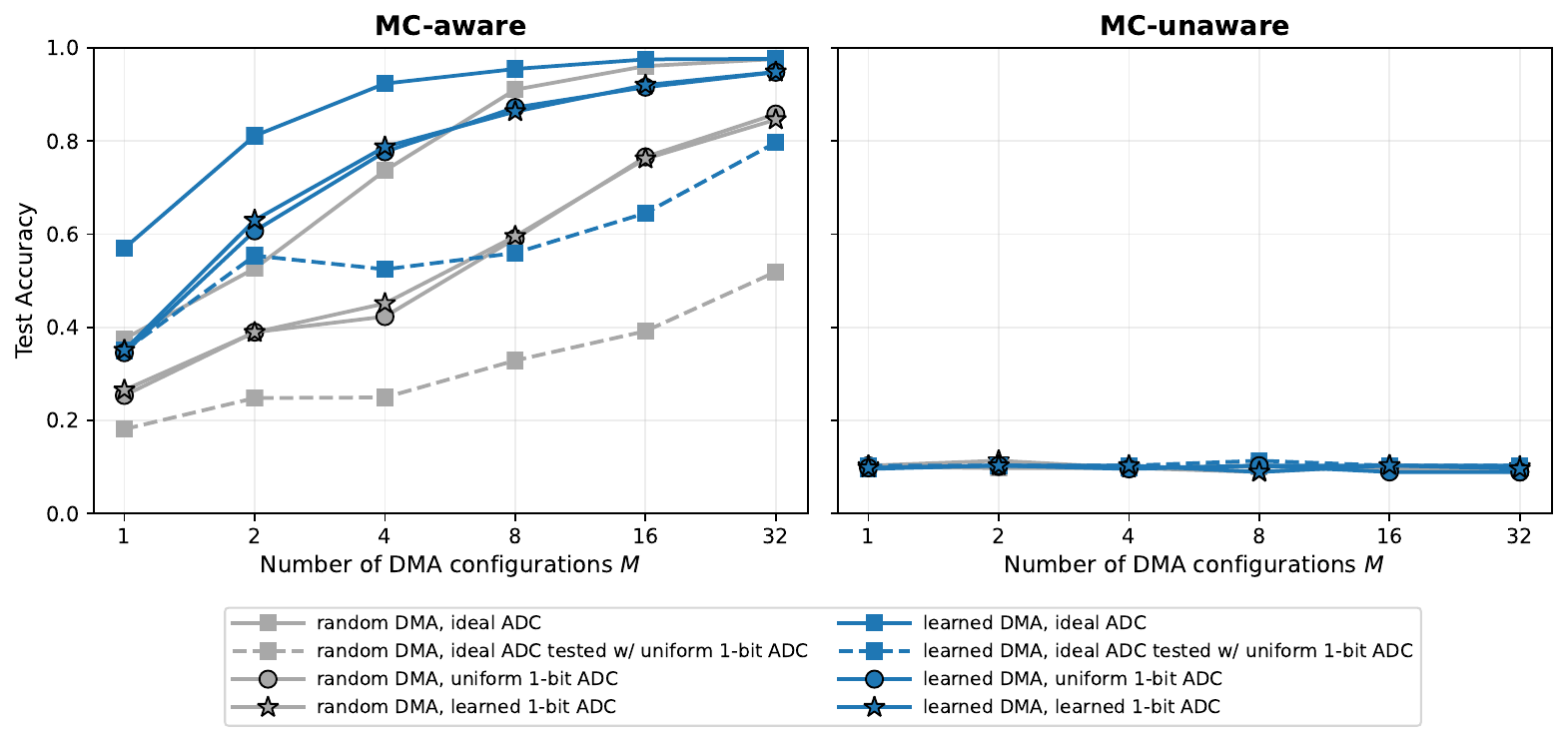}
    \caption{Scene-classification accuracy with one-bit ADCs as a function of the number of DMA configurations $M$, with (left) and without (right) MC awareness. The curves compare random (gray) and learned (blue) DMA configurations, for cases with ideal-ADC training (square symbol, solid line), deployment of ideal-ADC-trained models with a one-bit uniform ADC (square symbol, dashed line), ADC-aware training with a fixed uniform ADC (circle symbol), and joint DMA--ADC--digital training with learned ADC thresholds (star symbol).}
    \label{Fig4}
\end{figure*}

\subsubsection{Scene Classification with 1-Bit ADCs}
\label{subsubsec_Exp1bitClassifResults}

We begin with the maximally ADC-limited regime by setting $B=1$, such that each real-valued ADC branch has only two output levels. This case is practically relevant because one-bit ADCs are among the lowest-complexity and lowest-power quantization architectures, but it is also scientifically useful because it exposes the information loss caused by the ADC stage most clearly. We evaluate how the classification accuracy varies with the number of measurements $M$ and compare the different optimization cases introduced in Sec.~\ref{subsec_OptCases}. The key questions are: (i) how much performance is lost when ADC quantization is ignored during optimization (as in~\cite{del2020learned,qian2022noise}); (ii) how much of this loss can be recovered by ADC-aware DMA optimization with a fixed uniform ADC; and (iii) whether additionally learning the ADC thresholds in an end-to-end manner provides a further substantial gain under severe quantization.

Our results in the left panel in Fig.~\ref{Fig4} show that one-bit ADC quantization is a severe bottleneck unless it is explicitly accounted for during optimization. In the MC-aware case, learned DMA configurations optimized with an ideal ADC achieve high accuracies, increasing from $57.0\%$ at $M=1$ to $95.5\%$ at $M=8$. However, when the DMA configurations, ANN, and normalization parameters trained under the ideal-ADC assumption are kept fixed and the ideal ADC is replaced by a uniform one-bit ADC, the accuracy drops strongly; for example, at $M=8$ it decreases from $95.5\%$ to $56.0\%$. Thus, end-to-end optimization ignoring ADC quantization can yield a pipeline whose performance strongly deteriorates under ADC quantization.

Including the one-bit ADC during training recovers much of this loss. At $M=8$, ADC-aware optimization with a fixed uniform one-bit ADC improves the deployed one-bit accuracy from $56.0\%$ to $87.2\%$. The learned DMA configurations are also substantially more effective than random DMA configurations in the one-bit regime; for example, with a fixed uniform one-bit ADC and $M=8$, the learned-DMA accuracy is $87.2\%$, compared with $59.1\%$ for random DMA configurations. Learning the ADC thresholds provides at best marginal improvements: for the learned-DMA case, it improves the accuracy from $60.7\%$ to $63.0\%$ at $M=2$. This suggests that, in the one-bit regime considered here, the dominant benefit comes from ADC-aware task-specific DMA optimization, whereas learned non-uniform thresholds provide at most a secondary refinement.

Finally, the MC-unaware benchmark in the right panel of Fig.~\ref{Fig4} performs essentially at the random-guess level for all displayed values of $M$ and for all ADC settings. For instance, at $M=8$, the learned-DMA uniform-ADC accuracy drops from $87.2\%$ with MC-aware optimization to $10.3\%$ when MC is ignored during self-interference prediction and optimization. This confirms that, for the considered DMA prototype with strong MC, MC awareness is not a minor modeling refinement but a prerequisite for transferring optimized configurations to the MC-aware physical forward model.

\subsubsection{Scene Classification with Single-Shot Measurement}
\label{subsubsec_ExpSingleShotMeasurement}
We next consider the minimum-latency regime by setting $M=1$, assuming that latency is dominated by the number of sequentially used DMA configurations. The entire scene is then compressed into a single complex-valued microwave measurement. 

\begin{figure}
    \centering
    \includegraphics[width=\columnwidth]{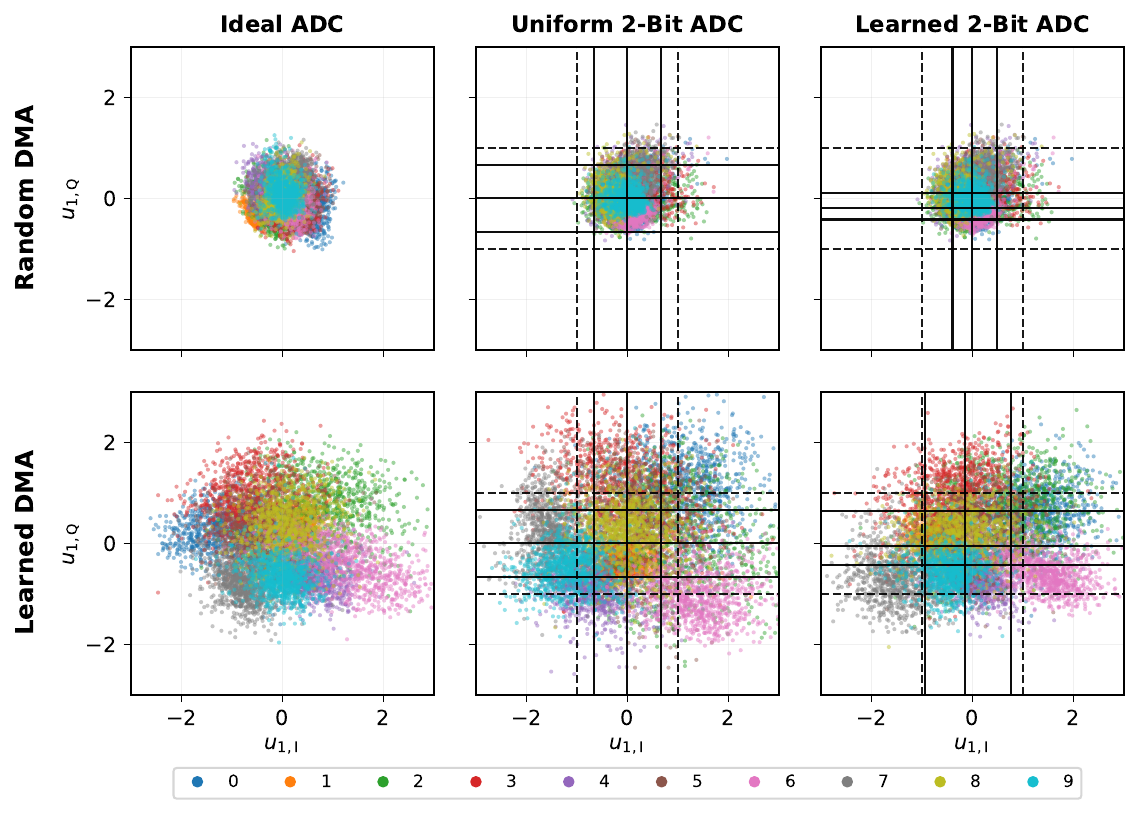}
    \caption{Test-set ADC inputs for $M=1$, shown in the normalized I-Q plane and colored by MNIST class. Rows compare random and learned DMA configurations; columns compare ideal, uniform 2-bit, and learned 2-bit ADCs. For finite-resolution ADCs, solid black lines mark internal bin boundaries and dashed black lines mark clipping boundaries. The \textit{random-DMA} distributions may differ across columns since they may be based on different random seeds.}
    \label{FigClouds}
\end{figure}

A useful feature of this single-shot measurement case is that the complex analog measurement vector $\mathbf z(\boldsymbol\rho,\mathbf V)$ reduces to a single scalar that can be visualized directly in the complex plane. In Fig.~\ref{FigClouds}, we compare \textit{random-DMA} and \textit{learned-DMA} scenarios with three ADC choices using the normalized ADC inputs $u_{m,\mathrm I}$ and $u_{m,\mathrm Q}$, for which the clipping boundaries are located at $-1$ and $+1$. The learned DMA configuration produces visibly more class-dependent structure than the random DMA configuration: the class-colored clouds are spread over a larger portion of the normalized I-Q plane and exhibit stronger clustering. This indicates that the DMA performs a task-aware wave-domain projection that pre-selects information useful for the downstream classification task. The learned DMA also uses the available analog dynamic range more aggressively, so that notably more samples exceed the clipping boundaries. Finally, the \textit{learned-ADC} cases visibly move the internal bin boundaries away from their uniform positions, with branch-dependent non-uniform spacings that adapt the quantization regions to the resulting signal distribution.

\begin{figure*}
    \centering
    \includegraphics[width=2\columnwidth]{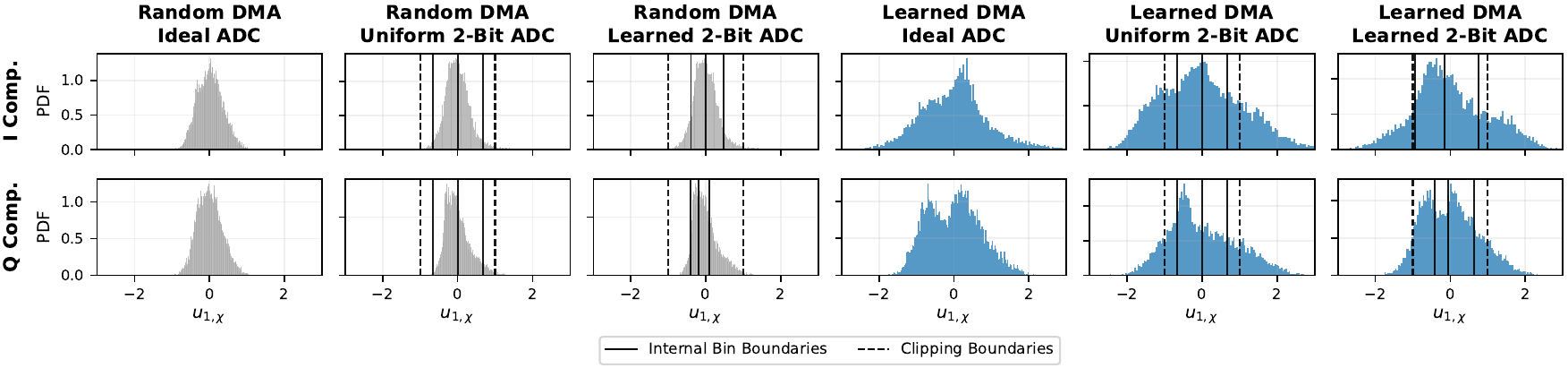}
    \caption{PDFs of the normalized I and Q ADC inputs corresponding to the test-set clouds shown in Fig.~\ref{FigClouds} for $M=1$. Rows show the I and Q components; columns compare random and learned DMA configurations with ideal, uniform 2-bit, and learned 2-bit ADCs. For finite-resolution ADCs, solid black lines mark internal bin boundaries and dashed black lines mark clipping boundaries. The \textit{random-DMA} distributions may differ across columns since they may be based on different random seeds.}
    \label{FigPDFs}
\end{figure*}

We show in Fig.~\ref{FigPDFs} the corresponding probability density functions (PDFs) of the I and Q components from Fig.~\ref{FigClouds}. The figure confirms that learned DMA configurations substantially reshape the analog measurement distribution, spreading it more broadly over the normalized ADC-input range than random DMA configurations. In the learned-ADC columns, the internal bin boundaries are notably placed non-uniformly and differently for the I and Q branches, indicating that the quantization regions adapt substantially to the signal statistics produced by the corresponding wave-domain front end.

\begin{figure}
    \centering
    \includegraphics[width=0.5\columnwidth]{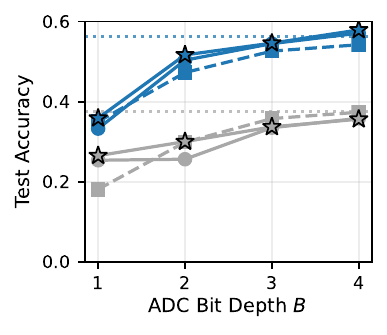}
    \caption{Scene-classification accuracy for the single-shot case $M=1$ as a function of ADC bit depth $B$, with MC awareness. Curve styles follow those explained in the caption of Fig.~\ref{Fig4}, with ``1-bit ADC'' replaced by ``$B$-bit ADC''. Dotted horizontal lines indicate the corresponding \textit{ideal-ADC} references.}
    \label{Fig8}
\end{figure}

In Fig.~\ref{Fig8}, we systematically evaluate the single-shot accuracy as a function of $B$. In the MC-aware case, \textit{learned-DMA} configurations consistently outperform \textit{random-DMA} configurations. With an \textit{ideal ADC}, the \textit{learned-DMA} case reaches $56.3\%$ test accuracy, compared with $37.6\%$ for the \textit{random-DMA} case. When the finite-resolution ADC is accounted for during training, the \textit{learned-DMA} accuracy increases from $33.3\%$ at $B=1$ to $57.1\%$ at $B=4$ for the \textit{uniform-ADC} case, and from $35.9\%$ to $57.9\%$ for the \textit{learned-ADC} case. Thus, for $M=1$, \textit{ADC-aware learned-DMA} optimization with only four ADC bits essentially recovers the \textit{ideal-ADC} performance; the slight excess over the \textit{ideal-ADC} reference should not be interpreted as a fundamental advantage of quantization, but rather as a consequence of separately optimized and validation-selected runs. The gain from learning the ADC thresholds is more modest than the gain from learning the DMA configurations: for the \textit{learned-DMA} case, the \textit{learned-ADC} curve improves over the \textit{uniform-ADC} curve by only $2.6$ percentage points for $B=1$ and $0.8$ percentage points for $B=4$.

\subsubsection{Ablation Studies for Measurement Noise Level}
\label{subsubsec_ablation_Noise}

Finally, we provide a brief ablation with respect to additive pre-ADC measurement noise. In contrast to the preceding results, which assume negligible measurement noise in order to isolate the effects of ADC awareness and MC awareness, we now train, validate and test each pipeline at the same prescribed noise level (i.e., our training is ``noise-aware''~\cite{qian2022noise}).
We generate the additive noise term \(n_m\) in \eqref{eq_single_measurement_si_cancelled} as zero-mean complex Gaussian noise whose real and imaginary parts are independent and have variance \(\sigma_n^2\). Hence, \(\mathbb E[|n_m|^2]=2\sigma_n^2\). The noise realization is redrawn whenever the measurement model is evaluated. We report the accuracy as a function of the absolute per-quadrature pre-ADC noise standard deviation \(\sigma_n\), rather than a normalized SNR, because the learned DMA configurations can themselves change the distribution of the absolute signal amplitudes.

\begin{figure}
    \centering
    \includegraphics[width=0.85\columnwidth]{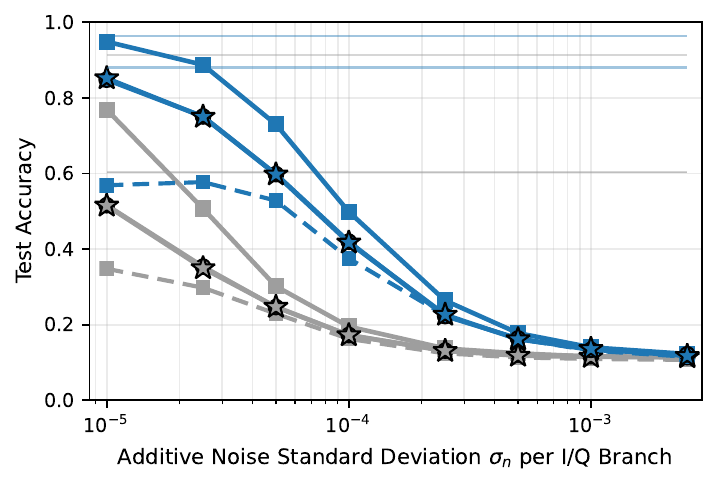}
    \caption{Pre-ADC measurement-noise ablation for the MC-aware case with \(M=8\) DMA configurations and 1-bit ADCs. Thin horizontal lines indicate the corresponding noiseless benchmarks. Curve styles follow those explained in the caption of Fig.~\ref{Fig4}.}
    \label{FigNoise}
\end{figure}

In Fig.~\ref{FigNoise}, we observe the expected accuracy degradation as the pre-ADC noise level increases. At low and moderate noise levels, learned DMA configurations remain clearly superior to random DMA configurations. For example, at \(\sigma_n=10^{-5}\), the learned-DMA uniform-ADC case reaches \(84.8\%\) accuracy, whereas the corresponding random-DMA case reaches only \(51.6\%\). At \(\sigma_n=10^{-4}\), the corresponding accuracies are \(42.0\%\) and \(17.3\%\), respectively. For larger noise levels, however, all curves approach the random-guess level of \(10\%\), indicating that the analog measurements themselves have become noise-limited.

The comparison between the learned-DMA ideal-ADC reference and the same ideal-ADC-trained system tested with a uniform one-bit ADC further illustrates the role of ADC awareness. At \(\sigma_n=10^{-5}\), this deployment mismatch reduces the accuracy from \(94.9\%\) to \(56.9\%\), whereas ADC-aware training with the fixed uniform one-bit ADC reaches \(84.8\%\). Hence, ADC awareness is still essential in the low-noise, low-resolution regime. As the noise level increases, this gap gradually vanishes because quantization mismatch ceases to be the dominant impairment. Finally, the learned-ADC and uniform-ADC curves remain almost identical throughout the sweep, confirming that, in the considered one-bit-ADC setting, most of the performance gain comes from task-aware DMA optimization and ADC-aware training rather than from learning non-uniform ADC thresholds.

\section{Conclusion}
\label{sec_Conclusion}

To summarize, we studied the ADC-aware end-to-end optimization of a DMA prototype with strong MC for monostatic scene classification. Our results show that ADC awareness is essential in low-resolution-ADC regimes: for instance, with $B=1$ and $M=8$, deploying an ideal-ADC-trained system with a uniform one-bit ADC reduces the accuracy from $95.5\%$ to $56.0\%$, whereas ADC-aware training with the same fixed uniform one-bit ADC recovers an accuracy of $87.2\%$. Ignoring MC during self-interference prediction and optimization deteriorates the accuracy to the random-guess level. Learned DMA configurations are particularly beneficial when the number of measurements is small; for $M=1$ with an ideal ADC, they improve the accuracy from $37.6\%$ for random DMA configurations to $56.3\%$. By contrast, learning non-uniform ADC thresholds provides only modest additional gains in the considered setup: for example, at $M=2$ and $B=1$, the accuracy increases only from $60.7\%$ with a fixed uniform ADC to $63.0\%$ with learned ADC thresholds.

Our separate sweeps over $M$ and $B$ further suggest that, under tight acquisition bit budgets, allocating bits to additional one-bit DMA measurements may be more beneficial than increasing the resolution of a single measurement; a systematic optimization over all $(M,B)$ pairs at fixed acquisition bit budget is left for future work. These findings contrast with the stronger learned-ADC gains reported in prior task-based ADC studies~\cite{shlezinger2022deep,vol2025learning}. 
A likely structural reason is that, in our pipeline, the dominant trainable acquisition degrees of freedom are the DMA configurations, while the learned ADC is a deliberately conservative threshold-only refinement with fixed reconstruction levels and shared I/Q scalar quantizers reused across the $M$ measurements. By contrast,~\cite{shlezinger2022deep} studies the optimization of a broader set of acquisition parameters including sampling choices, while~\cite{vol2025learning} focuses on trainable memristive ADC hardware and its power--accuracy trade-off. 
From a practical point of view, it is encouraging that for our DMA-based sensing pipeline most of the achievable performance gain can be reaped with ADC-aware wave-domain optimization and fixed uniform ADCs; indeed, this means that most of the observed performance gains can be obtained without custom trainable ADC hardware.


\bibliographystyle{IEEEtran}
\providecommand{\noopsort}[1]{}\providecommand{\singleletter}[1]{#1}%

\end{document}